\documentclass[aps,prl,reprint,
nofootinbib]{revtex4-2}
\usepackage{amsmath,amssymb,amsfonts}
\usepackage{bm}
\usepackage{graphicx}
\usepackage{xcolor}
\usepackage{hyperref}
\hypersetup{
    colorlinks=true,
    linkcolor=blue,
    citecolor=blue,
    urlcolor=blue
}

\usepackage{booktabs}
\usepackage{microtype}
\graphicspath{{./}{figures/}}
\newcommand{\D}{\mathcal D}

\newcommand{\ket}[1]{\lvert #1\rangle}
\newcommand{\bra}[1]{\langle #1\rvert}
\newcommand{\braket}[2]{\langle #1\vert #2\rangle}
\newcommand{\mean}[1]{\langle #1\rangle}
\newcommand{\rmd}{\mathrm d}

\begin{document}

\title{
Unraveling-Dependent Metastability in Monitored Quantum Systems and Associative Memories}

\author{Manali Malakar\textsuperscript{1}}
\author{Roberta Zambrini\textsuperscript{1}}
\author{Gian Luca Giorgi\textsuperscript{1}}
\affiliation{\textsuperscript{1}Institute for Cross-Disciplinary Physics and Complex Systems, IFISC (UIB-CSIC), Palma de Mallorca, Spain}

\date{\today}

\begin{abstract}
Metastability in open quantum systems is usually inferred from
spectral separation in the Liouvillian, which governs unconditional,
ensemble-averaged dynamics. We show that this diagnosis is incomplete
at the level of individual quantum trajectories: conditioned
realizations of the same unconditional dynamics can bypass,
transiently access, or operationally preserve a metastable memory,
depending on the monitored channel and the observed record. We
demonstrate these mechanisms in a driven-dissipative nonlinear
oscillator realizing a quantum associative memory, by comparing
spectra, target fidelities, and phase-space distributions. The
non-Hermitian dynamics obtained by post-selecting on the absence of
detected events supports metastable retrieval, but with a distinct
long-time fate: normalization selects the least-decaying mode of the
non-Hermitian spectrum rather than the addressed memory branch. In
contrast, stochastic jump trajectories can preserve retrieval over
extended times when the post-measurement update remains compatible
with the coherent memory structure. Thus trajectory-level
metastability is not determined by the averaged generator alone; it
requires compatibility between the monitored channel, the measurement
record, and the metastable manifold.
\end{abstract}

\maketitle

\emph{Introduction.---} 
Metastability in open quantum systems is conventionally diagnosed from the spectrum of the Liouvillian governing the ensemble-averaged density matrix. A separation between fast and slow decay modes implies that, after a short transient, the dynamics is confined to a low-dimensional metastable manifold before it eventually relaxes to the unique stationary state~\cite{Macieszczak2016,Macieszczak2021}. This spectral viewpoint has become a central tool for understanding long-lived behavior in driven-dissipative systems, including dissipative phase transitions, emergent classical dynamics~\cite{Macieszczak2021, Mignati2018}, synchronization~\cite{PhysRevLett.123.023604,Cabot2021}, and 
quantum associative memory, a mechanism by which a distorted input is dynamically corrected toward the nearest stored pattern via the slow manifold~\cite{rotondo2018,rebentrost2018quantum,fiorelli2020signatures,marsh2021enhancing,LabayMora2023,LabayMora2024,Labay-Mora_2025}.

However, the Liouvillian describes an average over all measurement records emitted into the environment. A continuously monitored system instead follows a quantum trajectory conditioned on the observed sequence of detection and no-detection events~\cite{Carmichael1993,Daley2014,wiseman_milburn}. The operational question is therefore not merely whether a slow manifold exists, but whether a given monitored realization retrieves, preserves, or erases a selected metastable state.

This question is especially natural for quantum associative memory, where retrieval must be meaningful in each individual
experimental run and not only after averaging.
A recent proposal showed that a single driven-dissipative nonlinear oscillator can realize such a memory by encoding the stored patterns as well-separated coherent-state lobes in phase space~\cite{LabayMora2023,LabayMora2024}. In the unconditional dynamics, nonlinear dissipation creates an $n$-branch memory manifold, while Liouvillian metastability allows distorted inputs to be retrieved during a long transient before the unique stationary state is reached. This establishes the ensemble-averaged memory mechanism, but leaves open whether the same metastable memory has a unique meaning at the trajectory level.

Recent work has analyzed how metastable open-system dynamics manifests in individual quantum trajectories~\cite{minganti2019,Brown2024, Xiang2026}. Separately, continuous-monitoring schemes can generate inequivalent unravelings of the same master equation~\cite{Chruscinski2022}, and unravelings can be distinguished through nonlinear trajectory-level observables~\cite{Pinol2024}.  Here we address a distinct, branch-resolved problem: which monitoring schemes and observed records retrieve and retain the stored branch addressed by a distorted input. 

\begin{figure}
\centering
\hspace{-4mm}
\includegraphics[width=0.8\linewidth]{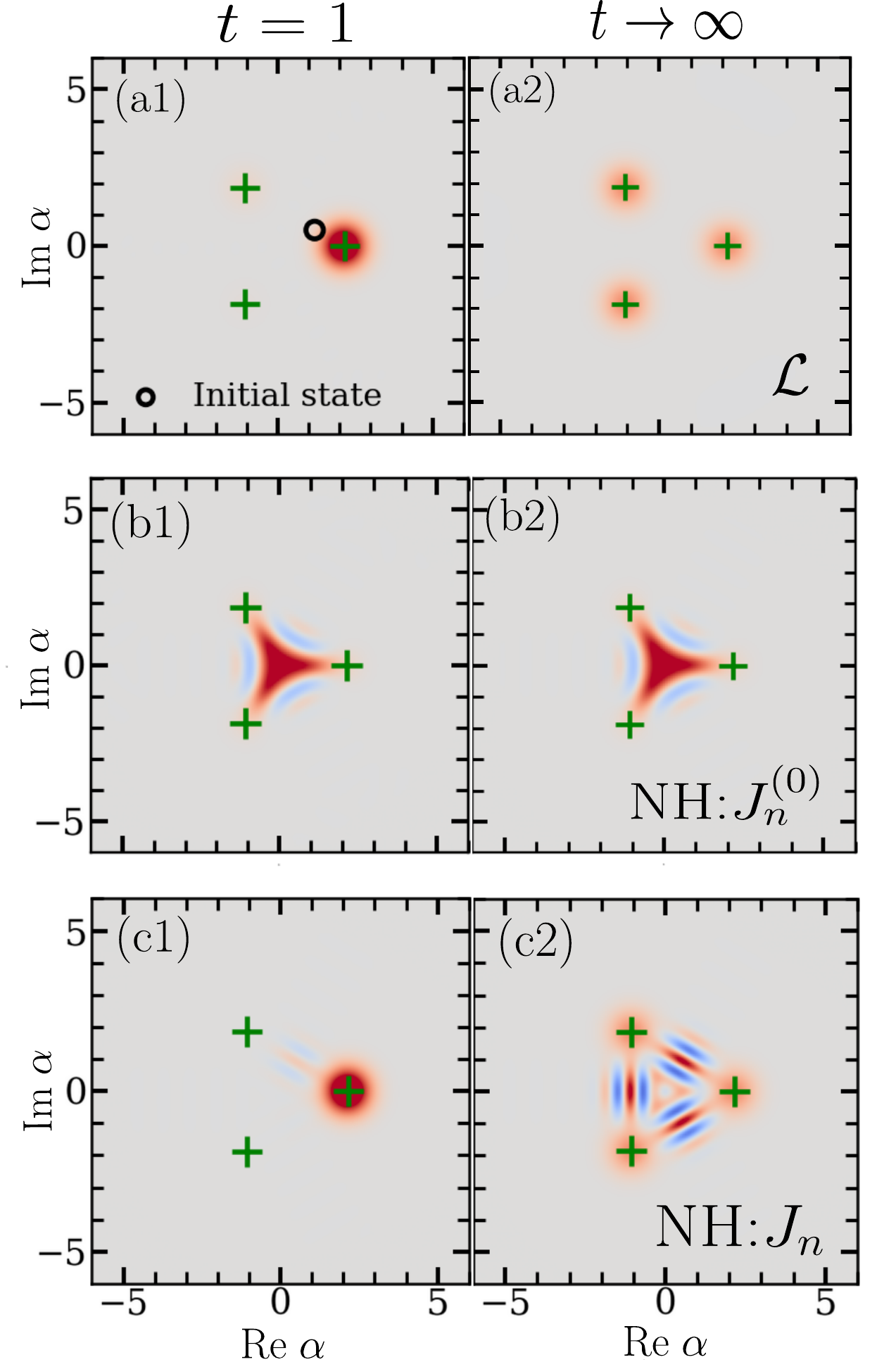}
\caption{Wigner distributions at $t=1$ (a1,b1,c1) and $t\to\infty$ (a2,b2,c2) for input $\alpha_{\mathrm{in}}=0.6\alpha_0e^{0.4i}$ (black circle). Panels (a1,a2) show Liouvillian dynamics $\mathcal L$; (b1,b2) no-click evolution with $J_n^{(0)}=\sqrt{\gamma_n}a^n$; and (c1,c2) no-click evolution with $J_n=\sqrt{\gamma_n}(a^n-\beta_n)$. Green crosses mark $\alpha_j^n=\beta_n$. Parameters are $n=3$, $\Delta=0.01$, $\gamma_n=1.2$, $\eta=6$, $\theta=0$, $N_{\mathrm{NH}}=100$, and $N_{\mathcal L}=40$.}
\label{fig1}
\end{figure}

We show that the same Liouvillian memory manifold can yield three operational fates. The input may bypass the stored manifold, reach the addressed branch only transiently, or retain it over an extended observation window. Two independent mechanisms decide which fate occurs. Access requires that the monitored channel be dark on the stored patterns, while retention requires that the detected jumps leave the addressed branch invariant, up to normalization and an irrelevant phase. Both requirements constrain the monitored operators and their post-measurement updates, not the Liouvillian spectrum, and are therefore invisible to the averaged generator: two schemes unraveling the same master equation may differ in whether they  display these mechanisms. A survival-selection process governs the asymptotic fate of a no-click record. Normalization progressively amplifies the slowest-decaying component of the non-Hermitian (NH) generator, so the conditioned state is ultimately selected by survival rather than by the pattern initially addressed. This has no counterpart in the trace-preserving Liouvillian dynamics, and it confines no-click retrieval to a finite window even when compatibility holds. The resulting physical principle is that Liouvillian spectral separation determines the availability and ensemble lifetime of a
metastable manifold, whereas the monitored channel and its measurement record determine whether a particular branch is accessed and retained in an individual realization.
\\~\\
\emph{Memory manifold and monitored channels.---}  
We start from the driven-dissipative nonlinear oscillator introduced as a quantum associative-memory platform~\cite{LabayMora2023,LabayMora2024}. Its unconditional evolution is described by the Gorini-Kossakowski-Sudarshan-Lindblad (GKSL) master equation~\cite{Lindblad1976, Gorini1976}
\begin{equation}
\dot\rho=-i[H_n,\rho]+\gamma_1{\cal D}[a]\rho+\gamma_m{\cal D}[a^m]\rho ,
\label{eqn1}
\end{equation}
with \({\cal D}[J]\rho=J\rho J^\dagger-\{J^\dagger J,\rho\}/2\) and
\(H_n=\Delta a^\dagger a+i\eta(e^{in\theta}a^{\dagger n}-e^{-in\theta}a^{n})\).
For \(m=n\), the nonlinear drive and \(n\)-photon damping can equivalently be combined into a displaced dissipator~\cite{LabayMora2023, Mirrahimi2014, Mundhada2017}
\begin{equation}
\dot\rho=-i\Delta[a^\dagger a,\rho]+\gamma_1{\cal D}[a]\rho
+\gamma_n{\cal D}[a^n-\beta_n]\rho,
\label{eqn2}
\end{equation}
where \(\beta_n=(2\eta/\gamma_n)e^{in\theta}\) and $n> 1$. This model possesses a weak $Z_n$ symmetry \cite{Albert2014}. The stored
patterns are the \(n\) coherent branches satisfying
\(\alpha_j^n=\beta_n\), i.e.
\(\alpha_j=\beta e^{i2\pi j/n}\), where \(\beta  \equiv |\beta_n|^{1/n} e^{i\theta}\).  Thus \(\eta/\gamma_n\)
sets the branch radius and \(\theta\) fixes the global orientation of the memory pattern. Unless stated otherwise, we set \(\gamma_1=1\) as the reference scale, with frequencies and decay rates scaled by \(\gamma_1\) and time by \(\gamma_1^{-1}\). The associative memory mechanism~\cite{LabayMora2023,LabayMora2024} is shown in Figs.~\ref{fig1}(a1) and~\ref{fig1}(a2). Under the unconditional Liouvillian dynamics $\mathcal{L}$, a distorted input is first driven toward the branch it addresses: at $t=1$, within the metastable regime, the Wigner function is localized on that branch [Fig.~\ref{fig1}(a1)].  At long times, the system relaxes to the symmetric stationary state, which retains the $n$-lobe geometry while losing the identity of the addressed branch [Fig.~\ref{fig1}(a2)].

Although Eqs.~(\ref{eqn1}) and~(\ref{eqn2}) generate exactly the same
unconditional density-matrix evolution, they correspond to
different monitoring schemes~\cite{Wiseman1993, Chruscinski2022, wiseman_milburn}. Operationally, Eq.~(\ref{eqn1}) corresponds to direct counting of the nonlinear output, whereas Eq.~(\ref{eqn2}) describes displaced counting, implemented by interference with a coherent reference before photodetection~\cite{wiseman_milburn}. In Eq.~(\ref{eqn1}), the nonlinear drive remains in the Hamiltonian, and the detected nonlinear emission is represented by the bare jump operator \(J^{(0)}_{n}=\sqrt{\gamma_n}a^n\). In Eq.~(\ref{eqn2}), the same drive is absorbed into the displaced jump operator \(J_n=\sqrt{\gamma_n}(a^n-\beta_n)\). This distinction is invisible to unconditional Liouvillian evolution but decisive for conditioned trajectories, which experience different no-click conditions and post-measurement updates. 

In the quantum-jump description~\cite{Dalibard1992, Plenio1998}, a click during an interval $dt$ occurs
with probability $\langle J_{\mu}^\dagger J_{\mu}\rangle dt$. For a coherent state, the detection rate associated with displaced nonlinear monitoring is  
 \(\langle J_n^\dagger J_n\rangle_{\alpha}=\gamma_n\lvert\alpha^n-\beta_n\rvert^2\). Its zeros coincide with the stored branches
$\alpha_j^n=\beta_n$. The absence of nonlinear detections
therefore filters the state toward the memory condition. By contrast, the bare \(n\)-photon loss rate \(\langle J_n^{(0)\dagger} J_n^{(0)}\rangle_{\alpha}=\gamma_n\lvert\alpha\rvert^{2n}\) is
minimized at the origin, and
$\ker(a^n)=\operatorname{span}
\{\lvert0\rangle,\ldots,\lvert n-1\rangle\}$ does not coincide
with the finite-radius memory manifold. The two monitoring
representations thus provide a direct test of whether a
Liouvillian metastable manifold remains operationally accessible
under conditioning.

\begin{figure}
\includegraphics[width=1.0\linewidth]{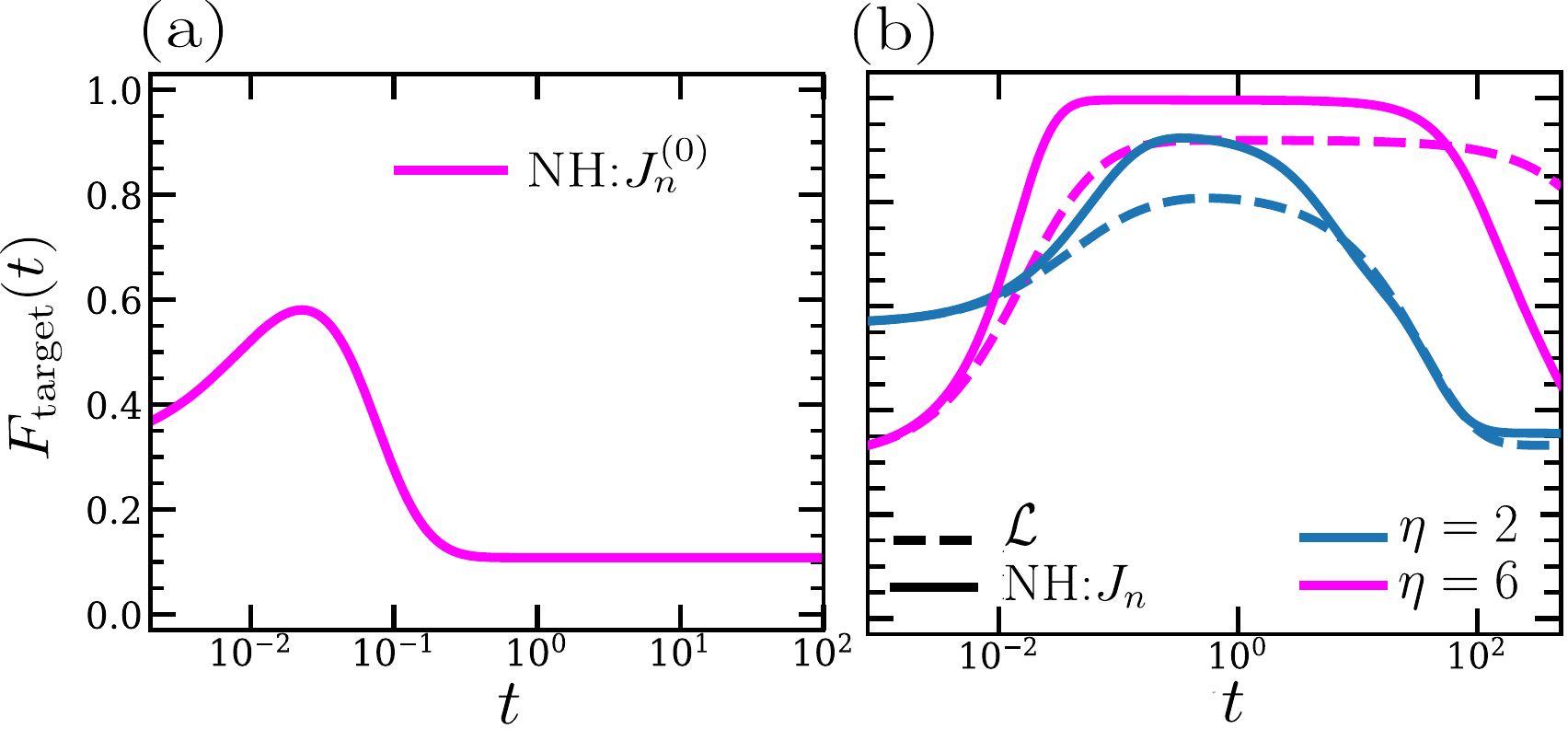}
\caption{Target fidelity $F_{\mathrm{target}}(t)$. (a) No-click evolution with the bare channel $J_n^{(0)}$. (b) No-click dynamics with $J_n$ (solid) and unconditional Liouvillian dynamics (dashed) for the indicated $\eta$. The initial state and remaining parameters are as in Fig.~\ref{fig1}.}
\label{fig2}
\end{figure}

Figs.~\ref{fig1}(b1) and~\ref{fig1}(b2) show the consequence of conditioning on the bare nonlinear channel. Even though the unconditional
Liouvillian supports a finite-radius memory manifold [Fig.~\ref{fig1}(a1)], the no-click state does not form a branch-localized lobe near the addressed pattern. The nearly identical Wigner distributions at
$t=1$ and at long times show that the trajectory rapidly
approaches the dominant non-Hermitian mode without passing
through the target branch. This is confirmed by the fidelity $F_{\mathrm{target}}(t)=\langle\alpha_0|\rho(t)|\alpha_0\rangle$, which reduces to $F_{\mathrm{target}}(t)=|\langle\alpha_0|\psi(t)\rangle|^2$ for pure states, between the targeted memory $\alpha_0$ and the evolved state. For the bare nonlinear channel, the fidelity [in Fig.~\ref{fig2}(a)] exhibits only a modest transient increase and remains far below unity. This is the first trajectory-level fate:
the memory manifold exists in the unconditional dynamics but is
bypassed by a monitoring condition whose dark structure is
incompatible with the stored patterns. Let us now consider the displaced memory channel $J_n$. The dynamics for no-click conditioning in Figs.~\ref{fig1}(c1) and~\ref{fig1}(c2) shows how the state retrieves the corresponding memory at finite time, whereas the long-time normalized state differs from all the others.
\\~\\

\emph{No-click condition and non-Hermitian dynamics.---}  
Let us analyze how the memory manifold is reached when no detections are observed in the monitored channels.  The no-click trajectory is generated by the effective non-Hermitian Hamiltonian
\(H_{\rm eff}=H-i(J_1^\dagger J_1+J_n^\dagger J_n)/2\)~\cite{Daley2014, Carmichael1993, Plenio1998, wiseman_milburn, ashida2020}.  For an unnormalized state \(|\tilde\psi(t)\rangle\), the norm
\(P_0(t)=\langle\tilde\psi(t)|\tilde\psi(t)\rangle\) is the probability of the no-click record, and the conditioned state is
\(|\psi(t)\rangle=|\tilde\psi(t)\rangle/\sqrt{P_0(t)}\).  This normalization is the physical selection imposed by the record: components with larger total click rate lose norm faster and are suppressed.  For the nonlinear memory channel, this has an interpretation in terms of retrieval. Since \(J_n|\alpha_j\rangle=0\) on the ideal memory branches \(\alpha_j^n=\beta_n\), the absence of nonlinear detections suppresses components away from \(a^n\simeq\beta_n\), and deviating states are conditionally filtered toward the memory manifold. This is the access condition of the
introduction, here satisfied by construction. 

The same conditioning makes no-click retrieval intrinsically finite-time. In the ideal nonlinear limit, $J_n=\sqrt{\gamma_n}(a^n-\beta_n)$ has an
$n$-dimensional dark kernel $\mathcal H_D=\ker(a^n-\beta_n)$, from which the
branch-localized memory states are formed. Ordinary one-photon loss is not dark on these branches, since $\langle J_1^\dagger J_1\rangle_{\alpha_j}
=\gamma_1|\alpha_j|^2$. Within \(\mathcal H_D\), its no-click contribution therefore lifts the degeneracy through weak mode dependence, producing small decay-rate splittings. Writing $H_{\mathrm{eff}}|R_k\rangle=E_k|R_k\rangle$ with
$E_k=\omega_k-i\Gamma_k$ and $\Gamma_0\le\Gamma_1\le\cdots$, the relative
amplitude of mode $k$ falls as $e^{-(\Gamma_k-\Gamma_0)t}$ because the
conditioned state is renormalized at every instant. A branch-localized superposition therefore survives only while these splittings remain unresolved; at longer times the conditioned state is selected by survival rather than by the branch addressed by the initial clue, and no-click retrieval is intrinsically finite-time.

The Wigner distributions in Figs.~\ref{fig1}(a1) and~\ref{fig1}(c1) contrast conditioned and unconditional evolution. At $t=1$, both are localized near the addressed branch and access the same metastable memory manifold, but by different mechanisms: no-click conditioning remains pure and filters components with large nonlinear detection rates, whereas Liouvillian evolution averages over jump outcomes and produces a mixed metastable state. At long times, the branch-localized no-click state [Fig.~\ref{fig1}(c1)] is lost [Fig.~\ref{fig1}(c2)] as survival selection leaves the least-decaying eigenmode of $H_{\mathrm{eff}}$. The trace-preserving Liouvillian instead redistributes weight among metastable branches through jump recycling. Its stationary Wigner function can therefore retain the $n$-lobe geometry [Fig.~\ref{fig1}(a2)] while erasing the initially addressed branch.

Fig.~\ref{fig2}(b) quantifies the retrieval dynamics. For both $\eta=2$ and $6$, no-click conditioning gives higher target fidelity because absence of nonlinear detections directly selects states satisfying $a^n\simeq\beta_n$. The Liouvillian plateau is lower because jump averaging dilutes the addressed component, but persists longer because of slow relaxation within the branch-population manifold. Increasing $\eta$ separates the coherent branches and improves retrieval in both descriptions, with no-click fidelity approaching unity for $\eta=6$. Importantly, $F_{\mathrm{target}}(t)$ is conditioned on observing no clicks, whereas $P_0(t)$ is the probability of that record. For $\eta=6$, $F_{\mathrm{target}}$ first reaches $0.99$ at $t_\star\simeq0.046$, when $P_0\simeq0.257$; at $t=1$, $F_{\mathrm{target}}\simeq0.996$ while $P_0\simeq0.003$. Thus retrieval begins before the selected record becomes rare. The later fidelity plateau describes the normalized state within a shrinking no-click subensemble, motivating the jump-resolved analysis below. The full $P_0(t)$ is shown in the Supplemental Material~\cite{SupplementalMaterial}.

The finite retrieval windows have a direct spectral origin. For $H_{\mathrm{eff}}$, the $n$ slowest modes form a non-Hermitian memory subspace separated from the fast spectrum: in the weak-detuning regime \(\Delta \ll \gamma_1\), coherent splittings are subleading, so the outer separation controls rapid entrance, whereas the smaller internal decay-rate splittings govern reweighting and eventual survival selection. The Liouvillian exhibits an analogous hierarchy, with rapid relaxation into the metastable manifold followed by much slower branch-population evolution. To substantiate
this spectral interpretation, in the End Matter we resolve
both spectra, define the associated entrance, reweighting,
and selection time scales, and identify their microscopic
origin through dynamics projected onto the nonlinear dark
manifold; derivations are given in the Supplemental Material. The spectral-separation ratios grow with $\eta$, showing that increasing branch separation widens the intermediate regime in which metastable retrieval is operational.
\\~\\
\emph{Metastable memory under stochastic jump trajectories.---} The no-click sector isolates a post-selected record in which all emissions are
absent. We now retain the detected events and evolve exact waiting-time
quantum-jump trajectories~\cite{Carmichael1993,Daley2014,wiseman_milburn,
Dalibard1992,Plenio1998}; algorithmic details are
given in the Supplemental Material
~\cite{SupplementalMaterial}. Between clicks, the state evolves under
$H_{\rm eff}$, while a detection in channel $\mu$ produces the normalized
update
\(|\psi_r\rangle\longrightarrow J_\mu|\psi_r\rangle/\|J_\mu|\psi_r\rangle\|.
\)
The two monitored channels $\mu \in \{1,n\}$ act differently within the memory manifold. On a stored
coherent branch $|\alpha_j\rangle$, with $\alpha_j^n=\beta^n$,
\begin{equation}\label{eqn9}
 \frac{J_1|\alpha_j\rangle}{\|J_1|\alpha_j\rangle\|}
 =e^{i\arg\alpha_j}|\alpha_j\rangle,
 \qquad J_n|\alpha_j\rangle=0.
\end{equation}
Thus, a linear-loss click records an emission without displacing or deforming the coherent lobe, whereas a nonlinear click occurs only if the state has developed weight away from the dark memory condition.  Eq.~(\ref{eqn9}) is the retention condition of the introduction in explicit form: the resolved loss jump maps the addressed branch onto itself, providing a trajectory-level mechanism for retaining it despite detected emissions, in the spirit of pointer-state selection \cite{zurek2003}.

\begin{figure}
\centering
\includegraphics[width=0.95\linewidth]{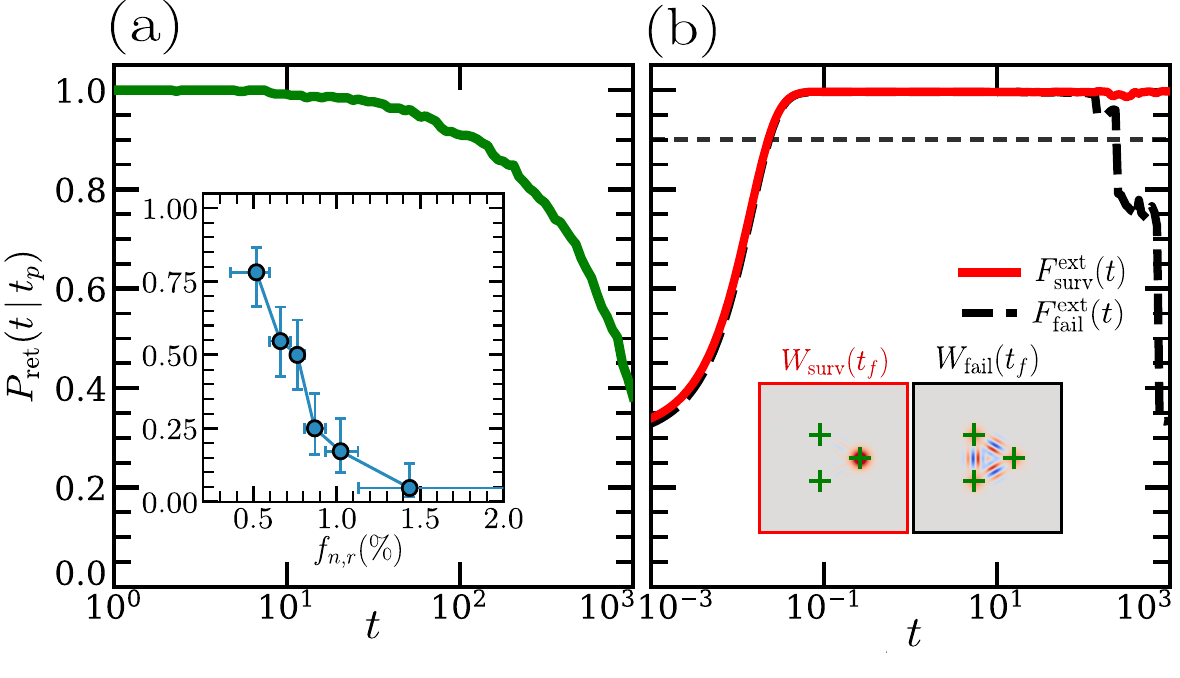}
\caption{\textit{Jump-resolved metastable retrieval}. (a) $P_{\rm ret}(t\mid t_p)$ for 500 exact waiting-time trajectories at $\eta=6$, conditioned on $F_r(t_p)\geq F_{\rm th}$ with $t_p=1$ and $F_{\rm th}=0.9$; 384 satisfy this condition. At $t_f=10^3$, 147 remain above threshold, giving $P_{\rm ret}(t_f\mid t_p)=147/384\simeq0.383$. The inset gives the endpoint probability in six equally populated bins of $f_{n,r}=N_{n,r}/(N_{1,r}+N_{n,r})$ over $(t_p,t_f]$; vertical bars are 95\% Wilson intervals~\cite{Wilson1927} and horizontal bars the observed bin ranges. (b) Extremal trajectories with the largest final fidelity above threshold and smallest below it; their Wigner functions at $t_f$ are shown below. Green crosses mark the memory roots, the gray dotted line $F_{\rm th}$. Times are in $\gamma_1^{-1}$; remaining parameters are as in Fig.~\ref{fig1}.}
\label{fig3}
\end{figure}

To quantify retrieval of the target $\alpha_{0}$ at a readout time $t$, we define the set of trajectories that have retrieved the target at the conditioning time $t_p$ as $\mathcal S(t_p)=\{r:F_r(t_p)\geq F_{\rm th}\}$, with $N_{\rm ret}(t_p)=|\mathcal S(t_p)|$, where
$F_r(t)=|\langle\alpha_0|\psi_r(t)\rangle|^2$. We take $t_p=1$ and $F_{\rm th}=0.9$. Their conditional instantaneous retrieval probability at a later readout time $t > t_p$ is
\begin{equation}
 P_{\rm ret}(t\mid t_p)=
 \frac{1}{N_{\rm ret}(t_p)}
 \sum_{r\in\mathcal S(t_p)}
 \Theta\!\left[F_r(t)-F_{\rm th}\right],
\end{equation}
where $\Theta$ is the Heaviside step function, with $\Theta(0)=1$. Fig.~\ref{fig3}(a) demonstrates that stochastic monitoring maintains a finite probability of retrieving the addressed branch over an extended observation window. Immediately after retrieval, $P_{\rm ret}(t\mid t_p)$ remains close to unity, showing that most conditioned trajectories are localized near the target branch at the queried time. At later times, it decreases as different jump records drive trajectories away from that branch. 

Crucially, $P_{\rm ret}(t_f\mid t_p)$ remains finite at $t_f=10^3$: the addressed memory is still localized and retrievable in a nonzero fraction of the conditioned trajectories, the operational signature of trajectory-level metastability. Because $P_{\rm ret}(t\mid t_p)$ is evaluated at the readout time, uninterrupted localization is unnecessary; trajectories that leave and return before $t$ are retained. The inset relates final retrieval to the measurement record. For each trajectory retrieved at $t_p$, $f_{n,r}=N_{n,r}/(N_{1,r}+N_{n,r})$ uses the linear and nonlinear counts over $(t_p,t_f]$. In six equally populated bins, $P_{\rm ret}(t_f\mid t_p)$ decreases with $f_{n,r}$, so records with more nonlinear detections are less likely to retrieve the addressed branch at the final readout. Since $J_n$ annihilates the ideal memory states, such detections signal excursions from the dark memory condition.

Fig.~\ref{fig3}(b) illustrates the two trajectory-level outcomes
underlying this ensemble behavior. To make their contrast visible, we show
extremal members of the ensemble: the surviving trajectory with
the largest final fidelity and the failed trajectory with the smallest final
fidelity. Both initially retrieve the addressed branch and follow nearly the
same high-fidelity plateau, but their late-time dynamics separate:
$F_{\rm surv}^{\rm ext}(t)$ remains close to unity, whereas
$F_{\rm fail}^{\rm ext}(t)$ undergoes jump-dependent excursions and ends
below $F_{\rm th}$. The Wigner functions at $t_f$ reveal that $W_{\rm surv}(t_f)$ remains localized at the addressed memory root, while $W_{\rm fail}(t_f)$ is distributed over phase space and has lost localization on that branch.

To test whether long-lived trajectory-level retrieval
depends on how the monitored channel acts on the memory
states, we replace the linear loss channel by linear gain as employed in quantum van der Pol self-oscillators~\cite{Lee2013}. This changes
the unconditional generator, so it is a control model rather than an
alternative unraveling. The comparison remains controlled: $J_n$ is unchanged,
so the stored patterns occupy the same phase-space locations and the
Liouvillian still supports a metastable manifold of the same $n$-branch
structure~\cite{Cabot2021}. Since
$a^\dagger|\alpha_j\rangle\not\propto|\alpha_j\rangle$, a gain click produces a
photon-added state and drives the trajectory away from the coherent branch.
The resulting trajectories exhibit only transient retrieval, as shown in the
Supplemental Material~\cite{SupplementalMaterial}. The two models differ in the detected jump of the linear channel, while their normalized no-click evolution is unchanged; retention is therefore determined by whether that jump preserves the coherent branch.
\\~\\
\emph{Conclusions.---} 
We have shown that Liouvillian spectral separation identifies an ensemble metastable manifold but does not determine its operational fate under continuous monitoring: two unravelings of the \emph{same} unconditional generator can yield qualitatively different trajectory-level outcomes. In the oscillator memory, a distorted input can bypass the memory manifold, reach the addressed branch only transiently, or retain it over an extended observation
window. These outcomes are set by two independent mechanisms.  Compatibility between the monitored channel and the metastable manifold controls whether the manifold is accessed and preserved: no-click conditioning retrieves the addressed branch only when the monitored dark condition coincides with the memory condition, and a jump-resolved record retains it only when the dominant post-measurement updates leave the coherent lobe invariant, as for photon loss, for which coherent states are pointer states, but not for the nonlinear channel or for a gain control. Survival selection, by contrast, controls the asymptotic fate within the no-click sector: normalization amplifies the slowest-decaying non-Hermitian mode, so conditioned retrieval is intrinsically finite-time, a mechanism absent from the trace-preserving Liouvillian dynamics. Compatibility thus decides whether
the memory is reached and the jump records preserve it; within the no-click sector, survival selection determines its asymptotic fate.

The measurement record is therefore an active component of associative retrieval, not a passive observation of ensemble relaxation. For a monitored quantum associative memory, the measurement channel is a design parameter that determines whether a stored pattern is genuinely retrieved in an individual run or merely present as a transient feature of the ensemble average. Beyond this system, the same trajectory-resolved framework can be extended to many-body metastable manifolds and to inefficient or diffusive detection; most directly, because compatibility fixes the trajectory-level fate, record-dependent feedback could steer the observed record toward compatible outcomes and stabilize retrieval in monitored quantum associative memories.\\~\\
\emph{Acknowledgements.---}We acknowledge support from the Spanish State Research Agency, through the María de Maeztu project CEX2021-001164-M, funded by MICIU/AEI/10.13039/501100011033; through the CoQuSy project PID2022-140506NB-C21 and -C22 funded by MICIU/AEI/10.13039/50110001103 and by ERDF, EU; and through the QuantCom project CNS2024-154720, funded by MICIU/AEI/10.13039/501100011033 and co-funded by the European Union; the project is funded under the Quantera II program that has received funding from the EU’s H2020 research and innovation program under Grant Agreement No. 101017733, and from the Spanish State Research Agency (project QNet PCI2024-153410) funded by MICIU/AEI/10.13039/50110001103  and by ERDF, EU.\\~\\
\emph{Data Availability.---}The data supporting the findings of this study are available from the authors upon reasonable request.

\bibliographystyle{apsrev4-2}
\bibliography{refs}

\onecolumngrid
\vspace{2.0em}
\begin{center}
{\Large\bfseries End Matter}
\end{center}
\vspace{1.0em}
\twocolumngrid

\setcounter{equation}{0}
\renewcommand{\theequation}{A\arabic{equation}}
\renewcommand{\theHequation}{A.\arabic{equation}}
\emph{Spectral structure and metastable time scales.---}
The relevant spectral scales characterizing relaxation into the
non-Hermitian memory subspace and the subsequent loss of branch
selectivity within it are defined as follows. Let $H_{\mathrm{eff}}|R_k\rangle=E_k|R_k\rangle$,
with $E_k = \omega_k - i\Gamma_k$ and $\Gamma_0 \le \Gamma_1 \le \cdots$. We
identify the $n$ slowest modes $k = 0,\ldots,n-1$ as the NH memory subspace.
Its separation from the remaining spectrum is set by the adjacent outer gap
\begin{equation}
\Delta^{\mathrm{NH}}_{\mathrm{out}} = \Gamma_n - \Gamma_{n-1},
\label{eqn3}
\end{equation}
while the internal splittings that reweight the slow modes are set by
\begin{equation}
  \tau^{\mathrm{NH}}_{\mathrm{rw}} = (\Gamma_{n-1} - \Gamma_0)^{-1}.
  \label{eqn4}
\end{equation}
\begin{figure}[h!]
\centering
\includegraphics[width=1.0\linewidth]{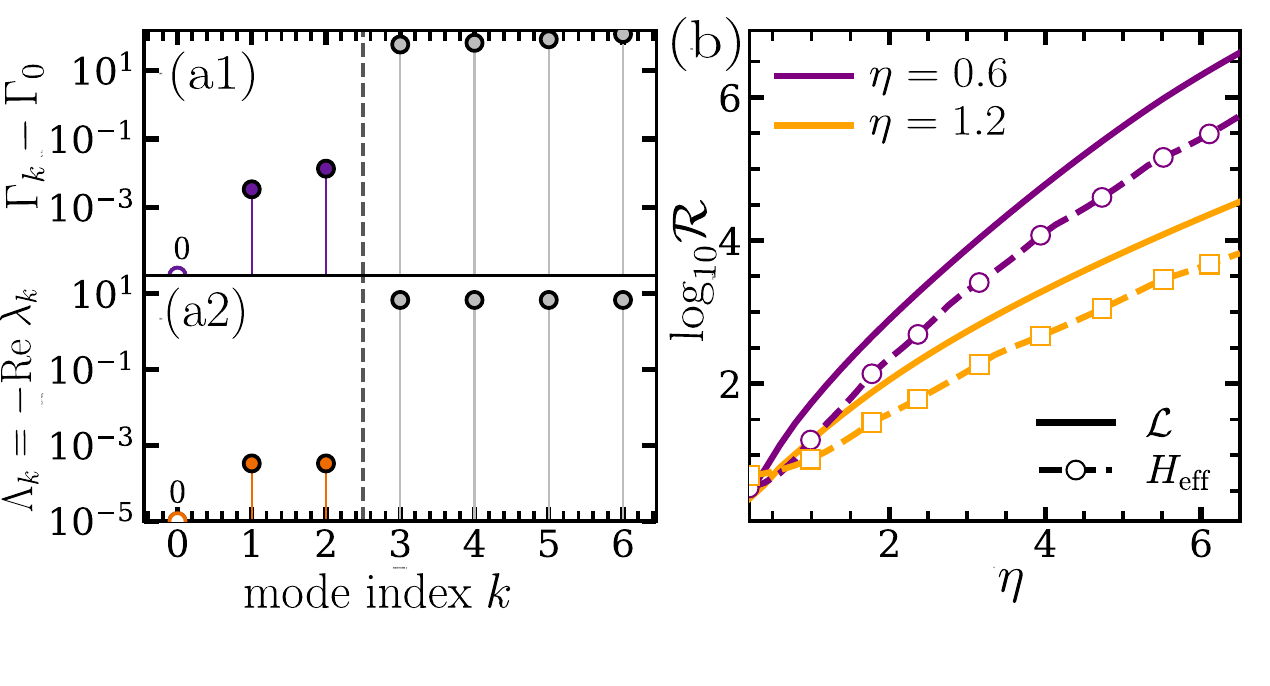}
\caption{Spectral separation of the slow subspaces for \(n=3\). (a1) Relative amplitude-decay rates \(\Gamma_k-\Gamma_0\) of \(H_{\mathrm{eff}}\) and (a2) Liouvillian decay rates \(\Lambda_k=-\mathrm{Re}\;\lambda_k\), both ordered from \(k=0\), at \(\eta=6\) and \(\gamma_n=1.2\). The purple points in (a1) and orange points in (a2) mark the three slow modes \(k=0,1,2\); gray points mark the fast sector, and the vertical dashed line separates the two. (b) \(\log_{10}\mathcal{R}_{\mathcal{L}}\) (solid) and \(\log_{10}\mathcal{R}_{\mathrm{NH}}\) (dashed with open markers) versus \(\eta\) for \(\gamma_n=0.6\) and \(1.2\). Their growth quantifies the widening time window between entrance into and evolution within the corresponding slow subspace. Throughout, \(\Delta/\gamma_1=0.01\), \(N_{\mathcal{L}}=40\), and \(N_{\mathrm{NH}}=100\).}
\label{fig_gap}
\end{figure}
The entrance time $\tau^{\mathrm{NH}}_{\mathrm{in}} = (\Gamma_n-\Gamma_0)^{-1}$
satisfies $(\tau^{\mathrm{NH}}_{\mathrm{in}})^{-1} =
\Delta^{\mathrm{NH}}_{\mathrm{out}} + (\tau^{\mathrm{NH}}_{\mathrm{rw}})^{-1}$
and therefore reduces to $1/\Delta^{\mathrm{NH}}_{\mathrm{out}}$ whenever the
outer gap dominates. The branch-preserving regime is then
\begin{equation}
\tau^{\mathrm{NH}}_{\mathrm{in}} \ll t \ll \tau^{\mathrm{NH}}_{\mathrm{rw}},
\label{eqn5}
\end{equation}
after which survival selection sets in, converging on the least-decaying mode
$|R_0\rangle$ on the scale $\tau^{\mathrm{NH}}_{\mathrm{sel}} =
(\Gamma_1-\Gamma_0)^{-1}$.
The resulting structure of the NH spectrum is shown in
Fig.~\ref{fig_gap}(a1). A local expansion of the nonlinear jump operator about a stored branch gives the large-\(|\beta|\) outer gap asymptotics
\begin{equation}
\Delta_{\mathrm{out}}^{\mathrm{NH}}=\frac{\gamma_n n^2}{2}|\beta|^{2n-2}\left[1-\frac{(n-1)^2}{2|\beta|^2}+O(|\beta|^{-4})\right].
\label{eqn6}
\end{equation}
The dark-state construction and the derivation of
Eq.~(\ref{eqn6}) are provided in the Supplemental Material
~\cite{SupplementalMaterial}. The internal NH splittings arise from the one-photon no-click term projected on the dark space. Let \(\{|\xi_q\rangle\}^{n-1}_{q=0}\) denote an orthonormal basis of the nonlinear dark manifold with projector \(P_{D}\). Within this manifold, $\Gamma_q\simeq\gamma_1\bar n_q/2$, where
$\bar n_q=\langle\xi_q|a^\dagger a|\xi_q\rangle$.
The weak sector dependence of $\bar n_q$, generated by
the residual overlap between the coherent branches,
produces the reweighting and final mode-selection scales.

For the full Liouvillian, let
$\mathcal L R_k=\lambda_kR_k$, with decay rates
$\Lambda_k=-\operatorname{Re}\lambda_k$ ordered as
$0=\Lambda_0\leq\Lambda_1\leq\cdots$. Here $R_k$ is an eigenmatrix in operator
space, distinguished by the absence of the ket from the Hilbert-space
eigenvectors $|R_k\rangle$ of $H_{\mathrm{eff}}$ introduced above.
A separation $\Lambda_{n-1}\ll\Lambda_{n}$ yields an
$n$-dimensional metastable manifold
~\cite{Macieszczak2016,Macieszczak2021}: the density
matrix enters this manifold on the timescale
$\Lambda_{n}^{-1}$, while its branch populations relax on the longer scale $\Lambda_{n-1}^{-1}$ [Fig.~\ref{fig_gap}(a2)]. The physical origin of this hierarchy follows by projecting the one-photon channel onto the nonlinear dark manifold~\cite{Kessler2012,ReiterSorensen2012}.
For well-separated branches,
$a_D=P_DaP_D\simeq\beta U$, where
$U|\xi_q\rangle=|\xi_{q-1}\rangle$, giving
\begin{equation}
\mathcal L_D^{(1)}\rho_D
\simeq
\gamma_1|\beta|^2
\left(U\rho_DU^\dagger-\rho_D\right).
\label{eqn7}
\end{equation}
This leading dynamics leaves the branch populations
stationary while damping interbranch coherences. Subleading
finite-$|\beta|$ corrections subsequently mix the
populations and generate the final relaxation to the unique
stationary state. The complete projected dynamics is
derived in the Supplemental Material
~\cite{SupplementalMaterial}.
\\~\\
To quantify the spectral separation in the two descriptions,
we define the following spectral ratios
\begin{equation}
\mathcal{R}_{\mathcal L}=\frac{\Lambda_{n}}{\Lambda_{n-1}},
\qquad
\mathcal{R}_{\mathrm{NH}}=\frac{\Gamma_n-\Gamma_0}{\Gamma_{n-1}-\Gamma_0}.
\label{eqn8}
\end{equation}
Fig.~\ref{fig_gap}(b) shows these ratios, whose growth with \(\eta\) yields a broader intermediate regime between entrance into and evolution within the respective slow subspaces.

\onecolumngrid
\clearpage
\setcounter{section}{0}
\setcounter{subsection}{0}
\setcounter{equation}{0}
\setcounter{figure}{0}
\setcounter{table}{0}
\renewcommand{\thesection}{S\arabic{section}}
\renewcommand{\thesubsection}{\thesection.\arabic{subsection}}
\renewcommand{\theequation}{S\arabic{equation}}
\renewcommand{\thefigure}{S\arabic{figure}}
\renewcommand{\thetable}{S\arabic{table}}
\renewcommand{\theHsection}{S.\arabic{section}}
\renewcommand{\theHsubsection}{\theHsection.\arabic{subsection}}
\renewcommand{\theHequation}{S.\arabic{equation}}
\renewcommand{\theHfigure}{S.\arabic{figure}}
\renewcommand{\theHtable}{S.\arabic{table}}
\begin{center}
{\Large\bfseries Supplemental Material for\\[0.4em]
``Unraveling-Dependent Metastability in Monitored Quantum Systems and Associative Memories''\par}
\vspace{1em}
Manali Malakar, Roberta Zambrini, and Gian Luca Giorgi
\end{center}
\vspace{1em}

This Supplemental Material establishes different measurement interpretations of the Lindblad representations, derives the non-Hermitian and Liouvillian slow
manifolds, quantifies the no-click probability, documents the numerical methods and convergence checks, and gives the linear-gain control discussed in the main text.  All dynamical results use the same distorted coherent
input \(\ket{\psi(0)}=\ket{\alpha_{\rm in}}=0.6\alpha_0 e^{0.4i}\),
and the addressed memory branch is $\ket{\alpha_0}$. Unless stated otherwise, we set $\gamma_1=1$ and use it as
the reference scale. The remaining parameters are
$n=3$, $\theta=0$, $\Delta=0.01$,
$\gamma_n=1.2$, and $\eta=6$. Thus all frequencies and
decay rates are scaled by $\gamma_1$, while time is measured
in units of $\gamma_1^{-1}$.

\section{Direct and displaced monitoring schemes}
\label{sec:gauge}

Throughout this work,
\begin{equation}
 \beta_n\equiv\frac{2\eta}{\gamma_n}e^{in\theta}
\label{eq:beta_definition}
\end{equation}
is a shorthand for the physical drive and nonlinear-damping parameters, not an
independent control parameter.  When an \(n\)th root is required below, we write
\(\beta^n=\beta_n\).  The algebraic equivalence of the two unconditional
representations follows from the single identity
\begin{equation}
 \D[A-\beta_n]\rho=\D[A]\rho
 -i\left[\frac{i}{2}
 \left(\beta_nA^\dagger-\beta_n^*A\right),\rho\right],
 \qquad A=a^n.
 \label{eq:gauge_identity}
\end{equation}
Multiplying Eq.~\eqref{eq:gauge_identity} by \(\gamma_n\) reproduces
\(H_n=i\eta(e^{in\theta}a^{\dagger n}-e^{-in\theta}a^n)\), and hence Eqs.~(1)
and (2) of the main text generate the same unconditional density matrix.
\begin{figure}[t]
 \centering
 \includegraphics[width=0.96\linewidth]{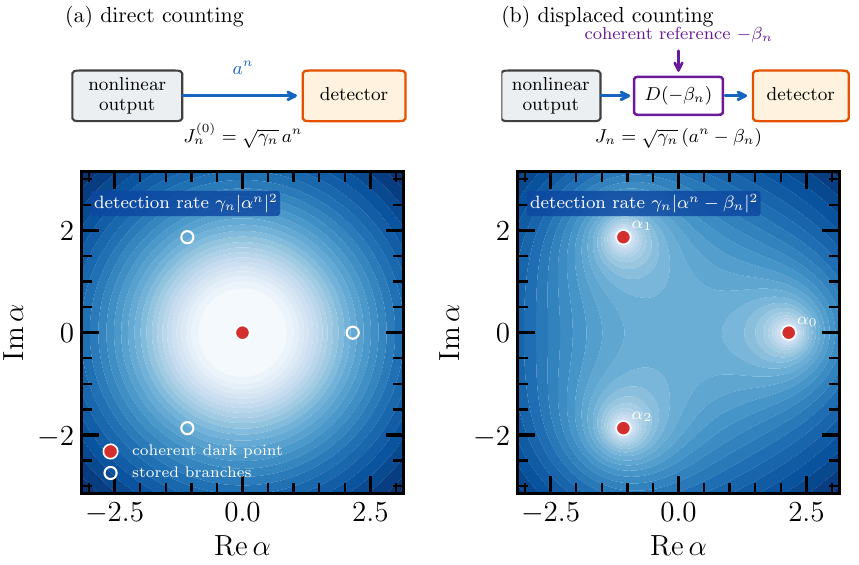}
 \caption{\label{fig:monitoring}
 Operational comparison of the monitored nonlinear channels.  The upper row
 shows direct counting and counting after coherent displacement \(D(-\beta_n)\) of the nonlinear output, implemented by interference with a coherent reference of normalized amplitude \(-\beta_n\) before the detector.  The lower row shows the corresponding semiclassical count rates for \(n=3\) and \(\eta=6\).  Direct counting has its coherent dark point at the origin, whereas displaced counting is dark at the three stored branches \(\alpha_j^n=\beta_n\) (red points).  The exact direct-channel quantum kernel is \(\operatorname{span}\{\ket{0},\ldots,\ket{n-1}\}\).  The two arrangements yield the same unconditional master equation but resolve different records.}
\end{figure}
Conditioned trajectories are nevertheless different because photodetection is
specified by the measured output operator.  Direct nonlinear counting uses
\begin{equation}
 J_n^{(0)}=\sqrt{\gamma_n}\,a^n,
 \qquad
 H^{(0)}=\Delta a^\dagger a+H_n,
\end{equation}
whereas displaced counting uses
\begin{equation}
 J_n=\sqrt{\gamma_n}\,(a^n-\beta_n),
 \qquad
 H=\Delta a^\dagger a.
\end{equation}
The displaced channel is implemented by interfering the nonlinear output with
a coherent reference before counting [\hyperlink{SM-WisemanMilburn2010}{1}], as illustrated
in Fig.~\ref{fig:monitoring}.  The corresponding no-click generators,
\begin{align}
 H_{\rm eff}^{(0)}&=H^{(0)}-\frac{i}{2}
 \left(J_1^\dagger J_1+J_n^{(0)\dagger}J_n^{(0)}\right),\\
 H_{\rm eff}&=H-\frac{i}{2}
 \left(J_1^\dagger J_1+J_n^\dagger J_n\right),
\end{align}
are different even though their jump-averaged master equations coincide.  Their
dark kernels are
\begin{align}
 \ker(a^n)&=\operatorname{span}\{\ket{0},\ldots,\ket{n-1}\},\\
 \ker(a^n-\beta_n)&=\operatorname{span}\{\ket{\xi_0},\ldots,
 \ket{\xi_{n-1}}\}.
\end{align}
Only the displaced channel is dark on the finite-radius branches
\(\alpha_j^n=\beta_n\).  This channel-dependent dark structure is the
operational origin of the direct-channel bypass and displaced-channel
retrieval shown in Figs.~1 and 2 of the main text.

\section{No-click probability and conditional fidelity}
\label{sec:p0}

\begin{figure}[t]
 \centering
 \includegraphics[width=0.96\linewidth]{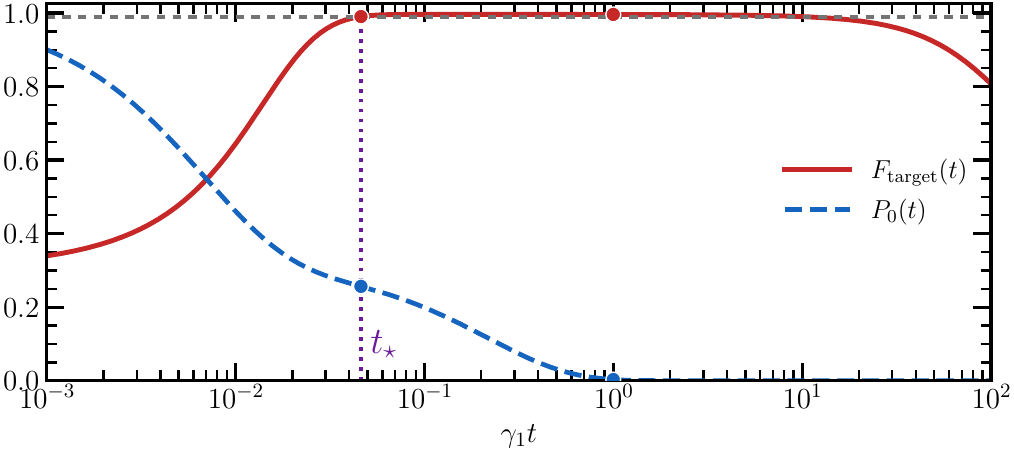}
 \caption{\label{fig:p0}
 Conditional target fidelity \(F_{\rm target}(t)\) (red solid line) and
 probability \(P_0(t)\) of the joint no-click record (blue dashed line) for the
 displaced channel at \(\eta=6\). The gray dashed horizontal line marks $F_{\mathrm{target}}=0.99$. Both quantities are shown on the
 same linear probability scale up to \(t=10^2\). The purple dotted
 line marks the first sampled time at which \(F_{\rm target}\geq0.99\), \(t_\star=0.04614\).  At this time
 \(F_{\rm target}=0.99054\) and \(P_0=0.2567\).  At \(t=1\),
 \(F_{\rm target}=0.99637\) and \(P_0=3.127\times10^{-3}\).
 The addressed branch is therefore retrieved before the selected record
 becomes rare.  Beyond \(t\sim1\), \(P_0\) is visually
 indistinguishable from zero on this linear scale, and the eventual fidelity
 decrease concerns an exceedingly small no-click subensemble.}
\end{figure}

For a pure initial state, the unnormalized no-click state is
\begin{equation}
 \ket{\widetilde\psi(t)}=e^{-iH_{\rm eff}t}\ket{\psi(0)},
 \qquad
 P_0(t)=\braket{\widetilde\psi(t)}{\widetilde\psi(t)}.
 \label{eq:p0_def}
\end{equation}
The normalized conditional state is
\(\ket{\psi_0(t)}=\ket{\widetilde\psi(t)}/\sqrt{P_0(t)}\).
Differentiating Eq.~\eqref{eq:p0_def} gives
\begin{equation}
 \frac{\rmd P_0}{\rmd t}
 =-P_0(t)\sum_\mu
 \bra{\psi_0(t)}J_\mu^\dagger J_\mu\ket{\psi_0(t)},
\end{equation}
where the sum runs over the monitored linear and nonlinear
channels, $\mu\in\{1,n\}$ and hence
\begin{equation}
 P_0(t)=\exp\left[-\int_0^t \rmd s\,
 h(s)\right],\qquad
 h(s)=\sum_\mu\mean{J_\mu^\dagger J_\mu}_{\psi_0(s)}.
 \label{eq:hazard}
\end{equation}
Thus \(P_0\) is the probability of the complete joint no-click record in the
linear and nonlinear monitored channels.  In contrast,
\begin{equation}
 F_{\rm target}(t)=\left|\braket{\alpha_0}{\psi_0(t)}\right|^2
\end{equation}
is a property of the normalized state conditional on that record.  A large
conditional fidelity is not a statement that the selected record is common.\\~\\
Let \(H_{\rm eff}\ket{R_k}=E_k\ket{R_k}\), with
\begin{equation}
 E_k=\omega_k-i\Gamma_k,\qquad \Gamma_k=-\operatorname{Im}E_k\geq0.
 \label{eq:gamma_convention}
\end{equation}
The convention in Eq.~\eqref{eq:gamma_convention} is an amplitude-decay
convention: an isolated eigenmode has amplitude \(e^{-\Gamma_k t}\) and
no-click probability \(e^{-2\Gamma_k t}\).  For a non-normal generator one
may use its biorthogonal right and left eigenvectors to write
\begin{equation}
 \ket{\widetilde\psi(t)}=
 \sum_k e^{-iE_kt}\ket{R_k}\braket{L_k}{\psi(0)}.
 \label{eq:biorthogonal}
\end{equation}
where the left and right eigenvectors are chosen biorthonormally,
\(\langle L_k|R_{k'}\rangle=\delta_{kk'}\). Eq.~\eqref{eq:biorthogonal} also shows why normalization ultimately
selects the least-decaying component with nonzero initial overlap.  The norm
contains interference terms when the right eigenvectors are not orthogonal,
so \(P_0(t)\) was evaluated directly from Eq.~\eqref{eq:p0_def}, not from an
incoherent sum of modal probabilities.\\~\\
Fig.~\ref{fig:p0} compares the quality and the probability of conditional retrieval for the displaced channel at $\eta = 6$.
Retrieval begins while the selected
record still has a probability of order $10^{-1}$: the fidelity first exceeds
$0.99$ at $t_\star = 0.046$, where $P_0 = 0.257$. By $t = 1$ the fidelity
remains high ($F_{\mathrm{target}} = 0.996$) but the record probability has
fallen to $3.1\times 10^{-3}$. 
The addressed branch is thus retrieved before
the no-click record becomes rare, while the eventual fidelity decrease at later times concerns an exceedingly small no-click subensemble. This limitation of post-selection is precisely what motivates the jump-resolved analysis of the main text, in which detection events are retained rather
than excluded.

\section{Non-Hermitian dark manifold and spectral time scales}
\label{sec:nh}

\subsection{Exact nonlinear dark states}

In the ideal nonlinear limit, $\Delta=\gamma_1=0$, the effective Hamiltonian reduces to
\begin{equation}
H_{\mathrm{eff}}^{\rm ideal}
=
-\frac{i\gamma_n}{2}
(a^{\dagger n}-\beta_n^*)
(a^n-\beta_n).
\label{eq:ideal_heff}
\end{equation}
A state is dark with respect to the nonlinear monitored
channel when
\begin{equation}
(a^n-\beta_n)\ket{\psi}=0,
\label{eq:dark_condition}
\end{equation}
or equivalently,
\begin{equation}
a^n\ket{\psi}
=
\beta_n\ket{\psi}.
\label{eq:dark_eigenvalue_condition}
\end{equation}
Such a state satisfies $J_n\ket{\psi}=0$ and therefore has
zero eigenvalue under $H_{\mathrm{eff}}^{\rm ideal}$.
\\
Since $a^n$ lowers the photon number by $n$, it preserves
the photon number modulo $n$. The Hilbert space consequently
decomposes into the mutually orthogonal residue sectors
\begin{equation}
\mathcal H_D
=
\bigoplus_{q=0}^{n-1}\mathcal H_q,
\qquad
\mathcal H_D
=
\operatorname{span}
\left\{
\ket{q+\ell n}:\ell=0,1,\ldots
\right\}.
\label{eq:mod_n_sectors}
\end{equation}
We seek one dark state in each sector in the form
\begin{equation}
\ket{\psi_q}
=
\sum_{\ell=0}^{\infty}
d_{\ell}^{(q)}
\ket{q+\ell n},
\qquad
q=0,\ldots,n-1.
\label{eq:general_sector_state}
\end{equation}
For $\ell\geq1$, the action of the $n$-fold annihilation
operator is
\begin{equation}
a^n\ket{q+\ell n}
=
\sqrt{
\frac{(q+\ell n)!}
     {[q+(\ell-1)n]!}
}
\ket{q+(\ell-1)n}.
\label{eq:an_fock_action}
\end{equation}
Applying $a^n$ to Eq.~\eqref{eq:general_sector_state}
therefore gives
\begin{align}
a^n\ket{\psi_q}
&=
\sum_{\ell=1}^{\infty}
d_{\ell}^{(q)}
\sqrt{
\frac{(q+\ell n)!}
     {[q+(\ell-1)n]!}
}
\ket{q+(\ell-1)n}
\nonumber\\
&=
\sum_{\ell=0}^{\infty}
d_{\ell+1}^{(q)}
\sqrt{
\frac{[q+(\ell+1)n]!}
     {(q+\ell n)!}
}
\ket{q+\ell n},
\label{eq:an_sector_state}
\end{align}
where the summation index was shifted in the second line.
On the other hand,
\begin{equation}
\beta_n\ket{\psi_q}
=
\sum_{\ell=0}^{\infty}
\beta_n d_{\ell}^{(q)}
\ket{q+\ell n}.
\end{equation}
Comparison of the coefficient of each
$\ket{q+\ell n}$ gives
\begin{equation}
d_{\ell+1}^{(q)}
\sqrt{
\frac{[q+(\ell+1)n]!}
     {(q+\ell n)!}
}
=
\beta_n d_{\ell}^{(q)},
\end{equation}
or equivalently,
\begin{equation}
d_{\ell+1}^{(q)}
=
\beta_n
\sqrt{
\frac{(q+\ell n)!}
     {[q+(\ell+1)n]!}
}
d_{\ell}^{(q)}.
\label{eq:dark_recurrence}
\end{equation}
For example, the first two coefficients are
\begin{align}
d_1^{(q)}
&=
\beta_n
\sqrt{\frac{q!}{(q+n)!}}\,
d_0^{(q)},
\nonumber\\
d_2^{(q)}
&=
\beta_n
\sqrt{\frac{(q+n)!}{(q+2n)!}}\,
d_1^{(q)}
\nonumber\\
&=
\beta_n^2
\sqrt{\frac{q!}{(q+2n)!}}\,
d_0^{(q)}.
\end{align}
Repeated application of the recurrence gives
\begin{align}
d_{\ell}^{(q)}
&=
\beta_n^\ell
\prod_{r=0}^{\ell-1}
\sqrt{
\frac{(q+rn)!}
     {[q+(r+1)n]!}
}
d_0^{(q)}
\nonumber\\
&=
\beta_n^\ell
\sqrt{
\frac{q!}{(q+\ell n)!}
}
d_0^{(q)}.
\label{eq:dark_coefficients}
\end{align}
The intermediate factorials cancel telescopically. Thus all
coefficients in sector $q$ are fixed by the single
coefficient $d_0^{(q)}$, which specifies the overall
normalization and phase.
\\
Substitution of Eq.~\eqref{eq:dark_coefficients} into
Eq.~\eqref{eq:general_sector_state} gives
\begin{equation}
\ket{\psi_q}
=
A_q
\sum_{\ell=0}^{\infty}
\frac{\beta_n^\ell}
     {\sqrt{(q+\ell n)!}}
\ket{q+\ell n},
\qquad
A_q\equiv\sqrt{q!}\,d_0^{(q)}.
\label{eq:unnormalized_dark_state}
\end{equation}
For the driven case $\beta_n\neq0$ considered here, choose
an $n$th root $\beta$ of $\beta_n$ and define
\begin{equation}
\beta_n=\beta^n,
\qquad
x=|\beta|^2=|\beta_n|^{2/n}.
\label{eq:beta_root_definition}
\end{equation}
We consider the normalization function
\begin{equation}
S_q(x)
=
\sum_{\ell=0}^{\infty}
\frac{x^{q+\ell n}}
     {(q+\ell n)!}.
\label{eq:dark_normalization}
\end{equation}
Using $|\beta_n|^{2\ell}=x^{\ell n}$, the norm of
Eq.~\eqref{eq:unnormalized_dark_state} is
\begin{align}
\braket{\psi_q}{\psi_q}
&=
|A_q|^2
\sum_{\ell=0}^{\infty}
\frac{x^{\ell n}}
     {(q+\ell n)!}
\nonumber\\
&=
\frac{|A_q|^2}{x^q}S_q(x).
\label{eq:unnormalized_dark_norm}
\end{align}
Normalization therefore fixes
\begin{equation}
|A_q|
=
\frac{x^{q/2}}{\sqrt{S_q(x)}}
=
\frac{|\beta|^q}{\sqrt{S_q(x)}}.
\end{equation}
The recurrence does not fix the overall phase of each
sector. We choose the convenient phase convention
\begin{equation}
A_q
=
\frac{\beta^q}{\sqrt{S_q(x)}}.
\label{eq:sector_phase_convention}
\end{equation}
Since $\beta_n^\ell=\beta^{\ell n}$, the normalized dark
state becomes
\begin{equation}
\ket{\xi_q}
=
\frac{1}{\sqrt{S_q(x)}}
\sum_{\ell=0}^{\infty}
\frac{\beta^{q+\ell n}}
     {\sqrt{(q+\ell n)!}}
\ket{q+\ell n},
\qquad
q=0,\ldots,n-1.
\label{eq:exact_dark_states}
\end{equation}
The factor $\beta^q$ in this expression is therefore not an
additional assumption. Its magnitude follows from
normalization, while its phase fixes the otherwise arbitrary
phase convention of sector $q$.
\\
The normalization is immediate:
\begin{equation}
\langle\xi_q|\xi_q\rangle
=
\frac{1}{S_q(x)}
\sum_{\ell=0}^{\infty}
\frac{x^{q+\ell n}}
     {(q+\ell n)!}
=
1.
\label{eq:dark_state_norm}
\end{equation}
The dark-state property can also be verified explicitly:
\begin{align}
a^n\ket{\xi_q}
&=
\frac{1}{\sqrt{S_q(x)}}
\sum_{\ell=1}^{\infty}
\frac{\beta^{q+\ell n}}
     {\sqrt{[q+(\ell-1)n]!}}
\ket{q+(\ell-1)n}
\nonumber\\
&=
\frac{1}{\sqrt{S_q(x)}}
\sum_{\ell=0}^{\infty}
\frac{\beta^{q+(\ell+1)n}}
     {\sqrt{(q+\ell n)!}}
\ket{q+\ell n}
\nonumber\\
&=
\frac{\beta^n}{\sqrt{S_q(x)}}
\sum_{\ell=0}^{\infty}
\frac{\beta^{q+\ell n}}
     {\sqrt{(q+\ell n)!}}
\ket{q+\ell n}
\nonumber\\
&=
\beta_n\ket{\xi_q}.
\label{eq:dark_state_verification}
\end{align}
Consequently,
\begin{equation}
(a^n-\beta_n)\ket{\xi_q}=0,
\qquad
J_n\ket{\xi_q}=0.
\label{eq:nonlinear_dark_property}
\end{equation}
For $q\neq q'$, the states $\ket{\xi_q}$ and
$\ket{\xi_{q'}}$ have support on disjoint sets of Fock
states and are therefore orthogonal:
\begin{equation}
\langle\xi_{q'}|\xi_q\rangle
=\delta_{q'q}.
\label{eq:dark_state_orthogonality}
\end{equation}
Moreover, Eq.~\eqref{eq:dark_recurrence} leaves exactly one
independent solution in each of the $n$ residue sectors.
Hence
\begin{equation}
\ker(a^n-\beta_n)
=
\operatorname{span}
\left\{
\ket{\xi_0},\ldots,\ket{\xi_{n-1}}
\right\},
\label{eq:exact_dark_kernel}
\end{equation}
and the ideal nonlinear dark manifold is exactly
$n$-dimensional. We denote the Hilbert-space projector onto the nonlinear
dark manifold by
\begin{equation}
P_D
=
\sum_{q=0}^{n-1}
\ket{\xi_q}\bra{\xi_q}.
\label{eq:dark_projector}
\end{equation}
The physical memory patterns are the branch-localized
coherent states
\begin{equation}
\alpha_j
=
\beta
\exp\left(\frac{2\pi i j}{n}\right),
\qquad
j=0,\ldots,n-1.
\label{eq:coherent_branches}
\end{equation}
Since $a\ket{\alpha_j}=\alpha_j\ket{\alpha_j}$ and
$\alpha_j^n=\beta^n=\beta_n$, each branch satisfies
\begin{equation}
(a^n-\beta_n)\ket{\alpha_j}=0.
\label{eq:coherent_dark_condition}
\end{equation}
Thus every coherent branch belongs to
$\ker(a^n-\beta_n)$. The $n$ amplitudes $\alpha_j$ are
distinct, and the corresponding coherent states are
linearly independent. Since the dark kernel is exactly
$n$ dimensional, $\{\ket{\alpha_j}\}_{j=0}^{n-1}$ forms a
nonorthogonal, branch-localized basis of the same manifold
spanned orthonormally by
$\{\ket{\xi_q}\}_{q=0}^{n-1}$.

\subsection{Lifting of the nonlinear degeneracy}
When \(\Delta=\gamma_1=0\), every state in Eq.~\eqref{eq:exact_dark_states} is dark and has
zero eigenvalue.  Projecting the detuning and the linear no-click loss onto
this manifold gives, to first order,
\begin{equation}
 P_DH_{\rm eff}P_D\simeq
 \left(\Delta-\frac{i\gamma_1}{2}\right)
 \sum_{q=0}^{n-1}\bar n_q\ket{\xi_q}\bra{\xi_q},
 \label{eq:projected_heff}
\end{equation}
where
\begin{equation}
 \bar n_q=\bra{\xi_q}a^\dagger a\ket{\xi_q}
 =x\frac{S_{q-1}(x)}{S_q(x)}.
 \label{eq:nbar}
\end{equation}
The subscript \(q-1\) in Eq.~\eqref{eq:nbar} is understood modulo \(n\).
Consequently,
\begin{equation}
 \omega_q\simeq\Delta\bar n_q,\qquad
 \Gamma_q\simeq\frac{\gamma_1}{2}\bar n_q.
\end{equation}
Therefore, the ratio of coherent to dissipative internal splittings is \(2\Delta/\gamma_1\), so coherent splittings are subleading in the weak-detuning regime \(\Delta\ll\gamma_1\). The weak sector dependence of \(\bar n_q\) comes from the exponentially small
overlap between different coherent branches. This sector dependence produces the internal
decay-rate bandwidth that eventually reweights a branch-localized
superposition.
\\~\\
Order the full non-Hermitian decay rates as
\(\Gamma_0\leq\Gamma_1\leq\cdots\). The $n$ slowest modes,
$\Gamma_0,\ldots,\Gamma_{n-1}$, define the
non-Hermitian memory subspace. We separate its internal
decay-rate bandwidth from the adjacent gap to the first
mode outside this subspace:
\begin{equation}
W_{\rm NH}
=
\Gamma_{n-1}-\Gamma_0,
\qquad
\Delta_{\rm out}^{\rm NH}
=
\Gamma_n-\Gamma_{n-1}.
\label{eq:nh_gap_bandwidth}
\end{equation}
These quantities satisfy
\(\Gamma_n-\Gamma_0=W_{\rm NH}+\Delta_{\rm out}^{\rm NH}\). Using the definition of $\mathcal{R}_{\rm NH}$ given in the End Matter, the spectral ratio plotted in Fig.~4 can therefore be
written as
\begin{equation}
\mathcal{R}_{\rm NH}
=
\frac{\Gamma_n-\Gamma_0}
     {\Gamma_{n-1}-\Gamma_0}
=
1+
\frac{\Delta_{\rm out}^{\rm NH}}
     {W_{\rm NH}}.
\label{eq:nh_spectral_ratio}
\end{equation}
Thus $\mathcal{R}_{\rm NH}\gg1$ corresponds to a separation between
the $n$-mode memory subspace and the faster modes that is
large compared with the decay-rate variation within the
memory subspace.

\subsection{Finite-radius correction to the non-Hermitian outer gap}

\begin{figure}[t]
\centering
\includegraphics[width=0.96\linewidth]{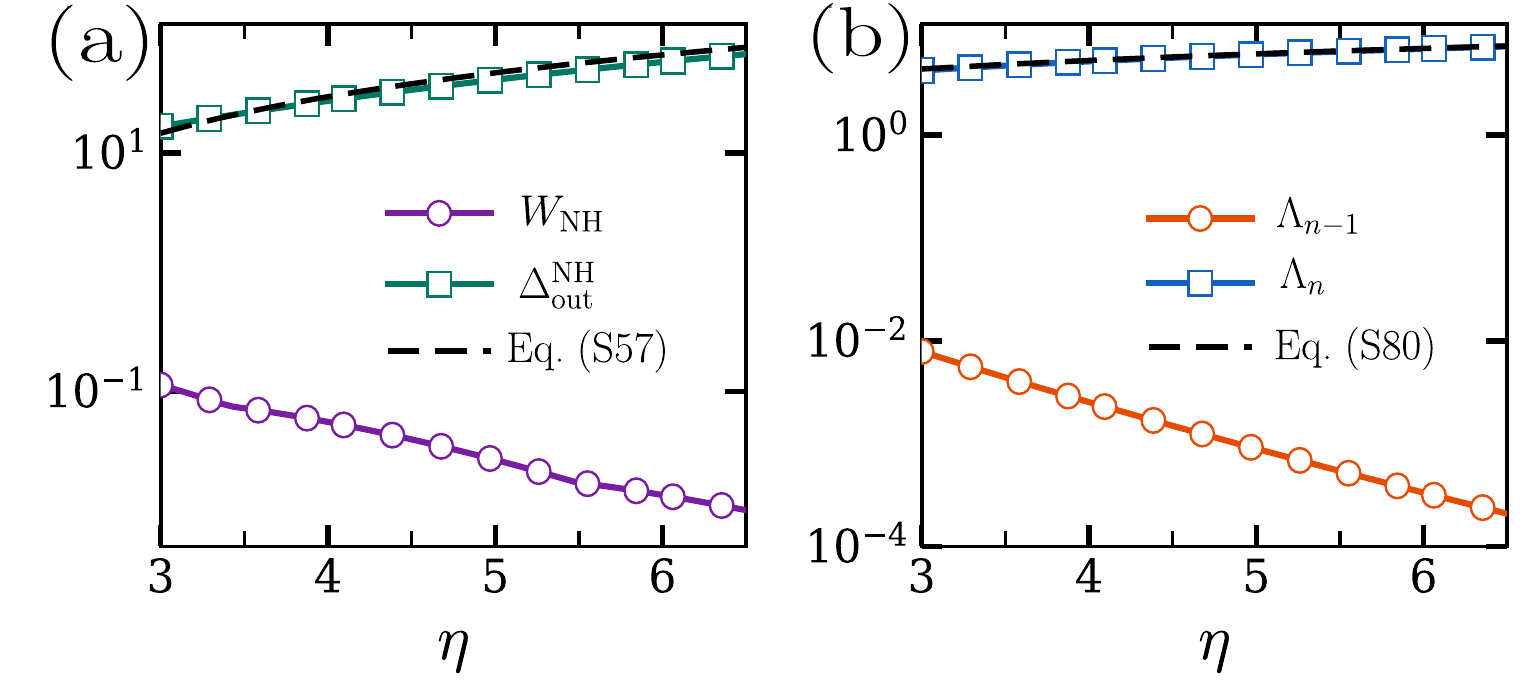}
\caption{\label{figS3}
\textit{Dependence of the non-Hermitian and Liouvillian
spectral scales on nonlinear drive.}
(a) Exact non-Hermitian internal bandwidth
$W_{\rm NH}=\Gamma_{n-1}-\Gamma_0$ and outer gap
$\Delta_{\rm out}^{\rm NH}=\Gamma_n-\Gamma_{n-1}$.
(b) Exact Liouvillian upper slow-sector rate $\Lambda_{n-1}$ and
first fast rate $\Lambda_n$, using the ordering
$0=\Lambda_0\leq\Lambda_1\leq\cdots$.
The black dashed curves show the analytical predictions for large-\(|\beta|\) of
Eq.~(\ref{eq:outer_correction}) for $\Delta_{\rm out}^{\rm NH}$ and Eq.~(\ref{eq:lambda_fast_large_radius}) for
$\Lambda_n$, respectively. Symbols denote independently calculated
numerical values, while solid lines guide the eye.
Parameters are $n=3$, $\gamma_n=1.2$, $\Delta=0.01$,
$N_{\rm NH}=100$, and $N_{\mathcal L}=40$.}
\end{figure}

To obtain the leading finite-radius correction, we expand
about a stored branch $\alpha_j$, satisfying
$\alpha_j^n=\beta_n$, by writing $a=\alpha_j+b$. Then
\begin{align}
a^n-\beta_n
&=n\alpha_j^{n-1}b+\frac{n(n-1)}{2}\alpha_j^{n-2}b^2+\cdots \nonumber\\
&=c_1 b+c_2 b^2+O(b^3),
\label{eq:local_jump_expansion}
\end{align}
where
\begin{equation}
c_1=n\alpha_j^{n-1},
\qquad
c_2=\frac{n(n-1)}{2}\alpha_j^{n-2}.
\label{eq:c1_c2}
\end{equation}
The nonlinear jump contributes the anti-Hermitian term
\begin{align}
H_{\rm eff}^{(n)}
&=
-\frac{i\gamma_n}{2}Q_j,
\nonumber\\
Q_j
&\equiv
(a^n-\beta_n)^\dagger(a^n-\beta_n)
\nonumber\\
&=
|c_1|^2b^\dagger b
+c_2^*c_1b^{\dagger 2}b
+c_1^*c_2b^\dagger b^2
+|c_2|^2b^{\dagger 2}b^2
+\cdots .
\label{eq:local_decay_operator}
\end{align}
At leading order,
$Q_j^{(0)}=|c_1|^2b^\dagger b$, whose eigenvalues in the
local fluctuation basis are
\begin{equation}
q_m^{(0)}=m|c_1|^2.
\end{equation}
Thus the first nonzero local eigenvalue is
$q_1^{(0)}=|c_1|^2$. The $n$ local vacua associated with
the stored branches span the $n$-dimensional nonlinear dark
manifold and generate the global slow sector
$\Gamma_0,\ldots,\Gamma_{n-1}$ when the weaker terms in
$H_{\rm eff}$ are restored. The first local fluctuation
sector is associated with the next group of modes, beginning
at $\Gamma_n$. Neglecting the much smaller decay-rate
bandwidth within the slow sector and the subleading effects
of linear loss and detuning, the outer gap is therefore
estimated as
\begin{equation}
\Delta_{\rm out}^{\rm NH}
\equiv
\Gamma_n-\Gamma_{n-1}
\simeq
\frac{\gamma_n}{2}q_1,
\label{eq:local_outer_gap}
\end{equation}
while its leading linearized value is
\begin{equation}
\Delta_{\rm out}^{(0)}
=
\frac{\gamma_n}{2}|c_1|^2
=
\frac{\gamma_n n^2}{2}|\beta|^{2n-2}.
\label{eq:outer_leading}
\end{equation}
Consequently, within this local large-$|\beta|$ estimate,
the relative correction to $q_1$ gives the same relative
correction to the outer gap.\\~\\
The correction terms displayed in
Eq.~\eqref{eq:local_decay_operator} have vanishing diagonal
matrix elements in $|1\rangle_b$: the cross terms change the
fluctuation number by one, while
$b^2|1\rangle_b=0$ eliminates the
$|c_2|^2b^{\dagger2}b^2$ term. The leading shift therefore
arises from second-order mixing between $|1\rangle_b$ and
$|2\rangle_b$.
\begin{equation}
{}_b\langle2|
c_2^*c_1 b^{\dagger 2}b
|1\rangle_b
=
\sqrt{2}\,c_1c_2^* .
\end{equation}
Second-order perturbation then gives
\begin{align}
\delta q_1
&=
\frac{
2|c_1|^2|c_2|^2
}{
q_1^{(0)}-q_2^{(0)}
}
\nonumber\\
&=
\frac{
2|c_1|^2|c_2|^2
}{
|c_1|^2-2|c_1|^2
}
=
-2|c_2|^2 .
\label{eq:first_eigenvalue_shift}
\end{align}
Consequently,
\begin{align}
q_1
&=
|c_1|^2
\left[
1-2\left|\frac{c_2}{c_1}\right|^2
+O(|\alpha_j|^{-4})
\right]
\nonumber\\
&=
|c_1|^2
\left[
1-\frac{(n-1)^2}{2|\alpha_j|^2}
+O(|\alpha_j|^{-4})
\right],
\label{eq:first_local_eigenvalue}
\end{align}
where
\begin{equation}
\frac{c_2}{c_1}
=
\frac{n-1}{2\alpha_j}.
\end{equation}
Terms involving $O(b^3)$ also have no diagonal matrix
element in $|1\rangle_b$ and first contribute at relative
order $|\alpha_j|^{-4}$. Hence no additional correction of
relative order $|\alpha_j|^{-2}$ is omitted. Writing
$\alpha_j=\beta e^{2\pi i j/n}$, with
$|\alpha_j|=|\beta|$, we finally obtain
\begin{equation}
\Delta_{\rm out}^{\rm NH}=
\Delta_{\rm out}^{(0)}
\left[
1-\frac{(n-1)^2}{2|\beta|^2}
+O(|\beta|^{-4})
\right].
\label{eq:outer_correction}
\end{equation}
Since $\Delta_{\rm out}^{(0)}\propto|\beta|^{2n-2}$,
Eq.~(\ref{eq:outer_correction}) predicts an increasing non-Hermitian outer gap, with a
relative finite-radius reduction
$(n-1)^2/(2|\beta|^2)$ that vanishes as $|\beta|\rightarrow\infty$.
Fig.~\ref{figS3}(a) shows that increasing the drive simultaneously enlarges $\Delta_{\rm out}^{\rm NH}$ and narrows the internal bandwidth $W_{\rm NH}$, thereby increasing the separation between the memory
subspace and the faster modes. The dashed curve evaluates Eq.~(\ref{eq:outer_correction})
over the displayed interval and captures the drive dependence of the
outer gap. The exact numerical rates are used for all quantitative
time windows, with the amplitude-decay convention of Eq.~(\ref{eq:gamma_convention}).

\section{Projected Liouvillian slow manifold}
\label{sec:projected_liouvillian}

In the ideal nonlinear limit, the Liouvillian is
\begin{equation}
\mathcal L_0\rho
=
\gamma_n\mathcal D[a^n-\beta_n]\rho .
\label{eq:ideal_nonlinear_liouvillian}
\end{equation}
Since
\begin{equation}
(a^n-\beta_n)\ket{\xi_q}=0
\end{equation}
for every $q=0,\ldots,n-1$, the $n$ dark kets generate the
$n^2$-dimensional dark operator space
\begin{equation}
\mathcal K_D
=
\operatorname{span}
\left\{
\ket{\xi_q}\bra{\xi_p}:
q,p=0,\ldots,n-1
\right\}.
\label{eq:dark_operator_space}
\end{equation}
The Hilbert-space projector $P_D$ was defined in
Eq.~\eqref{eq:dark_projector}. The corresponding
operator-space projection is
\begin{equation}
\mathcal P_D[X]=P_DXP_D.
\label{eq:dark_superprojector}
\end{equation}
For every operator
$\rho_D=P_D\rho_DP_D$ in $\mathcal K_D$, \(\mathcal L_0\rho_D=0\). All residue-sector indices below are understood modulo $n$.

\subsection{Projected one-photon operators}

The action of one-photon annihilation closes exactly within
the nonlinear dark-ket manifold. Using
Eq.~\eqref{eq:exact_dark_states},
\begin{align}
a\ket{\xi_q}
&=
\frac{1}{\sqrt{S_q(x)}}
\sum_{\ell=0}^{\infty}
\frac{
\beta^{q+\ell n}\sqrt{q+\ell n}
}{
\sqrt{(q+\ell n)!}
}
\ket{q+\ell n-1}
\nonumber\\
&=
\beta
\sqrt{
\frac{S_{q-1}(x)}{S_q(x)}
}
\ket{\xi_{q-1}}\equiv c_q\ket{\xi_{q-1}},
\label{eq:a_dark_action}
\end{align}
where
\begin{equation}
c_q
=
\beta
\sqrt{
\frac{S_{q-1}(x)}{S_q(x)}
}.
\label{eq:cq_definition}
\end{equation}
For $q=0$, the vacuum contribution vanishes, and shifting the remaining summation index gives the same result with $q-1\equiv n-1$. The coefficients satisfy
\begin{equation}
|c_q|^2
=
x\frac{S_{q-1}(x)}{S_q(x)}
=
\bar n_q,
\label{eq:cq_nbar_relation}
\end{equation}
where $\bar n_q$ is the mean occupation. The projected annihilation and number operators are
therefore
\begin{align}
a_D
&\equiv
P_DaP_D
=
\sum_{q=0}^{n-1}
c_q
\ket{\xi_{q-1}}\bra{\xi_q},
\label{eq:projected_a}
\\
N_D
&\equiv
P_Da^\dagger aP_D
=
a_D^\dagger a_D
\nonumber\\
&=
\sum_{q=0}^{n-1}
|c_q|^2
\ket{\xi_q}\bra{\xi_q}.
\label{eq:projected_number}
\end{align}
The equality
$P_Da^\dagger aP_D=a_D^\dagger a_D$ follows because
$a$ maps the dark-ket manifold exactly into itself.

\subsection{First-order effective Liouvillian}

The weak detuning and one-photon loss define the
perturbation
\begin{equation}
\mathcal V\rho
=
-i\Delta[a^\dagger a,\rho]
+
\gamma_1\mathcal D[a]\rho .
\label{eq:weak_liouvillian}
\end{equation}
Because $(a^n-\beta_n)P_D=0$, the nonlinear dissipator
vanishes identically inside the dark operator space: \(\mathcal D[a^n-\beta_n]\rho_D=0\), where
\(\rho_D=P_D\rho_DP_D\). Treating $\mathcal V$ perturbatively relative to the
spectral separation from the bright nonlinear modes, the first-order slow generator is
[\hyperlink{SM-Kessler2012}{5},\,\hyperlink{SM-ReiterSorensen2012}{6}]
\begin{align}
\mathcal L_D^{(1)}
&=
\mathcal P_D\mathcal V\mathcal P_D
=
-i\Delta[N_D,\,\cdot\,]
+
\gamma_1\mathcal D[a_D].
\label{eq:projected_liouvillian}
\end{align}
Higher-order terms contain virtual excursions into the
bright nonlinear subspace and are not included in
Eq.~\eqref{eq:projected_liouvillian}.
\\
To display the projected generator explicitly, introduce
the operator basis
\begin{equation}
R_{qp}^{(0)}
=
\ket{\xi_q}\bra{\xi_p},
\qquad
q,p=0,\ldots,n-1.
\label{eq:dark_operator_basis}
\end{equation}
Equations~\eqref{eq:projected_a} and
\eqref{eq:projected_number} give
\begin{align}
N_DR_{qp}^{(0)}
&=
|c_q|^2R_{qp}^{(0)},
\nonumber\\
R_{qp}^{(0)}N_D
&=
|c_p|^2R_{qp}^{(0)},
\nonumber\\
a_DR_{qp}^{(0)}a_D^\dagger
&=
c_qc_p^*
R_{q-1,p-1}^{(0)}.
\label{eq:projected_operator_actions}
\end{align}
Substituting these relations into the commutator and
dissipator yields
\begin{align}
\mathcal L_D^{(1)}R_{qp}^{(0)}
={}&
-i\Delta
\left(
|c_q|^2-|c_p|^2
\right)
R_{qp}^{(0)}-\frac{\gamma_1}{2}\left(|c_q|^2+|c_p|^2\right)
R_{qp}^{(0)}+\gamma_1c_qc_p^*R_{q-1,p-1}^{(0)}.
\label{eq:projected_liouvillian_matrix}
\end{align}
Eq.~\eqref{eq:projected_liouvillian_matrix} is the
explicit $n^2$-dimensional projected representation of the
slow Liouvillian. The first term describes the
sector-dependent coherent phase evolution, the second is
the anticommutator loss, and the final term is the
jump-recycling contribution that shifts both residue
indices. The finite-radius dependence within this first-order projection is retained through the coefficients \(c_q\).

\subsection{Large-radius limit and branch coherences}

For well-separated coherent branches,
\begin{equation}
S_q(x)
\simeq
\frac{e^x}{n}
\end{equation}
independently of $q$, up to corrections that are
exponentially small in the branch separation. Consequently,
\begin{equation}
c_q\simeq\beta,
\qquad
a_D\simeq\beta U,
\qquad
N_D\simeq|\beta|^2I_D,
\label{eq:large_radius_operators}
\end{equation}
where \(U\) is the cyclic shift operator:
\(U\ket{\xi_q}=\ket{\xi_{q-1}}\), and \(I_D=P_D\).
The projected detuning commutator vanishes at this order, and Eq.~\eqref{eq:projected_liouvillian} reduces to
\begin{equation}
\mathcal L_D^{(1)}\rho_D
\simeq
\gamma_1|\beta|^2
\left(
U\rho_DU^\dagger-\rho_D
\right).
\label{eq:large_radius_liouvillian}
\end{equation}
Introduce the eigenstates of the cyclic shift,
\begin{equation}
\ket{\phi_j}
=
\frac{1}{\sqrt n}
\sum_{q=0}^{n-1}
\exp\left(
\frac{2\pi i jq}{n}
\right)
\ket{\xi_q},
\qquad
j=0,\ldots,n-1.
\label{eq:shift_eigenstates}
\end{equation}
They satisfy
\begin{equation}
U\ket{\phi_j}
=
\exp\left(
\frac{2\pi i j}{n}
\right)
\ket{\phi_j}.
\label{eq:shift_eigenvalues}
\end{equation}
Their relation to the physical coherent branches follows by grouping the coherent-state Fock expansion into residue
sectors:
\begin{equation}
\ket{\alpha_j}
=
e^{-x/2}
\sum_{q=0}^{n-1}
\sqrt{S_q(x)}
\exp\left(
\frac{2\pi i jq}{n}
\right)
\ket{\xi_q}.
\label{eq:coherent_dark_expansion}
\end{equation}
When the branches are well separated,
$S_q(x)\simeq e^x/n$, and therefore \(\ket{\alpha_j}\simeq\ket{\phi_j}\). Thus the eigenstates of the cyclic shift become the
branch-localized memory states in the large-radius limit. Applying Eq.~\eqref{eq:large_radius_liouvillian} to
$\ket{\phi_m}\bra{\phi_\ell}$ gives
\begin{align}
\mathcal L_D^{(1)}
\ket{\phi_m}\bra{\phi_\ell}
={}&
\gamma_1|\beta|^2
\left[
\exp\left(
\frac{2\pi i(m-\ell)}{n}
\right)-1
\right]
\ket{\phi_m}\bra{\phi_\ell}.
\label{eq:large_radius_eigenoperators}
\end{align}
The corresponding decay rates are
\begin{equation}
-\operatorname{Re}\lambda_{m\ell}^{(1)}
=
\gamma_1|\beta|^2
\left[
1-
\cos\left(
\frac{2\pi(m-\ell)}{n}
\right)
\right].
\label{eq:large_radius_decay_rates}
\end{equation}
For $m\neq\ell$, the operators
$\ket{\phi_m}\bra{\phi_\ell}$ represent interbranch
coherences and decay on a time scale of order
$(\gamma_1|\beta|^2)^{-1}$. For $m=\ell$, the eigenvalue
vanishes at this leading large-radius order, leaving an
$n$-dimensional branch-population sector. Jump recycling
therefore suppresses coherences between the branches
without immediately selecting a unique branch population.
\\~\\
With ordering
$0=\Lambda_0\leq\Lambda_1\leq\cdots$, the $n$ branch-population
modes occupy the indices $0,\ldots,n-1$ at leading large radius.
The first mode associated with interbranch-coherence decay is
therefore $\Lambda_n$. As shown in Fig.~\ref{figS3}, finite-radius effects make $\Lambda_{n-1}$ small but nonzero, whereas $\Lambda_n$ remains set by the leading coherence-decay scale and increases with the
drive. Their opposite trends generate the growing Liouvillian
spectral separation. For $n=3$, Eq.~(\ref{eq:large_radius_decay_rates}) gives
\begin{equation}
 \Lambda_n^{(1)}
 =\frac{3}{2}\gamma_{1}|\beta|^2
 =\frac{3}{2}\gamma_{1}|\beta_n|^{2/3}.
 \label{eq:lambda_fast_large_radius}
\end{equation}
The black dashed curve in Fig.~\ref{figS3}(b) shows this prediction throughout the displayed interval. 

\paragraph*{\textit{Numerical validation.---}}
The calculations use \(N_{\rm NH}=100\), \(N_{\mathcal L}=40\), and
\(N=100\) for the no-click, Liouvillian, and jump evolutions, respectively. Both the loss- and gain-channel ensembles
contain $N_{\rm traj}=500$ trajectories. Further checks are summarized in
Table~\ref{tab:convergence}.
\begin{table}[htbp]
\caption{\label{tab:convergence}Numerical convergence and validation checks. The
reported fidelity differences are maxima over the full plotted time grid.}
\vspace{5mm}
\small
\setlength{\tabcolsep}{7pt}
\begin{tabular}{@{}l p{0.46\textwidth} l@{}}
\toprule
check & parameters & result\\
\midrule
NH cutoff & \(N_{\rm NH}=100\) versus \(N_{\rm NH}=200\), \(\eta/\gamma_1=6\)
& \(\max_t|\Delta F|=3.0\times10^{-11}\)\\
Liouvillian cutoff &
\(N_{\mathcal L}=40\) versus \(50\), \(\eta/\gamma_1=2\)
& \(\max_t|\Delta F|=4.6\times10^{-10}\)\\
Liouvillian cutoff &
\(N_{\mathcal L}=40\) versus \(50\), \(\eta/\gamma_1=6\)
& \(\max_t|\Delta F|=2.6\times10^{-8}\)\\
Loss trajectories & population in the highest five Fock levels at \(t_f\)
& \(0\) up to numerical precision \\
Gain trajectories & population in the highest five Fock levels at \(t_f\)
& \(0\) up to numerical precision\\
\bottomrule
\end{tabular}
\end{table}

\section{Exact waiting-time trajectories and ensemble statistics}
\label{sec:trajectories}
\subsection{Waiting-time sampling}

Each Monte Carlo wave-function trajectory is generated using
an exact waiting-time construction
[\hyperlink{SM-Dalibard1992}{2},\,\hyperlink{SM-PlenioKnight1998}{3}].
Let $|\psi(t_0)\rangle$ denote the normalized trajectory
state immediately after a jump, or at the initial time.
During an interval with no detected event, the corresponding
unnormalized state is
\begin{equation}
|\widetilde{\psi}(\tau)\rangle
=
e^{-iH_{\mathrm{eff}}\tau}
|\psi(t_0)\rangle .
\label{eq:unnormalized_waiting_state}
\end{equation}
Its squared norm,
\begin{equation}
S(\tau)
=
\langle\widetilde{\psi}(\tau)|
\widetilde{\psi}(\tau)\rangle ,
\label{eq:survival_function}
\end{equation}
is the conditional probability that no jump occurs during
the interval $(t_0,t_0+\tau]$. Accordingly,
\begin{equation}
w(\tau)
=
-\frac{dS(\tau)}{d\tau}
=
\sum_{\mu}
\langle\widetilde{\psi}(\tau)|
J_\mu^\dagger J_\mu
|\widetilde{\psi}(\tau)\rangle
\label{eq:waiting_density}
\end{equation}
is the probability density for the next jump to occur after
a waiting time $\tau$.
\\
To sample this distribution, a uniform random threshold
$r\in(0,1)$ is drawn. The trajectory evolves without a
jump while $S(\tau)>r$, and the next jump occurs at the
first-jump waiting time $\tau_j$ satisfying
\begin{equation}
S(\tau_j)=r .
\label{eq:waiting_root}
\end{equation}
This follows from $\Pr(T>\tau)=S(\tau)$, where $T$ is the random waiting time.
\\~\\
Immediately before the jump, the normalized state is
\begin{equation}
|\psi^{-}\rangle
=
\frac{|\widetilde{\psi}(\tau)\rangle}
{\sqrt{S(\tau)}} .
\label{eq:prejump_state}
\end{equation}
Conditional on a jump occurring at that time, channel
$\mu$ is selected with probability
\begin{equation}
p_\mu
=
\frac{
\langle\psi^-|J_\mu^\dagger J_\mu|\psi^-\rangle
}{
\sum_\nu
\langle\psi^-|J_\nu^\dagger J_\nu|\psi^-\rangle
}.
\label{eq:channel_probability}
\end{equation}
The normalized post-jump state is then
\begin{equation}
|\psi(t_0+\tau)\rangle
=
\frac{J_\mu|\psi^-\rangle}
{\|J_\mu|\psi^-\rangle\|}.
\label{eq:postjump_state}
\end{equation}
A new random threshold is drawn after every jump and the
procedure is repeated until the final time, and if no further
threshold crossing occurs before $t_f$, the state is
propagated without a jump to $t_f$. The root of Eq.~(\ref{eq:waiting_root}) is determined with relative and absolute tolerances $10^{-11}$. The jump times are continuous random variables and are independent of the
logarithmic output grid, which is used only to record
observables. Repeating this construction generates the complete sequence of jump times, detected channels, and conditional states for
each trajectory. We use these records below to define the
retrieval-conditioned ensemble and the final-fidelity
distribution.

\subsection{Trajectory ensemble and retrieval statistics}

\begin{figure}[htbp]
 \centering
 \includegraphics[width=0.5\linewidth]{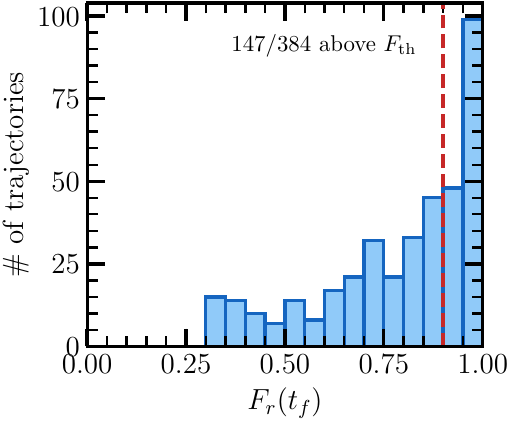}
 \caption{\label{fig:jump_stats}Final-fidelity distribution for the loss-channel trajectories.
The histogram shows $F_r(t_f)$ for all
$384$ trajectories in the retrieval-conditioned set
$\mathcal S(t_p)$. The dashed line marks
$F_{\mathrm{th}}=0.9$; $147$ trajectories lie above the
threshold, giving
$P_{\mathrm{ret}}(t_f|t_p)=0.3828$.
The corresponding 95\% Wilson score interval is
$[0.3356,0.4324]$.}
\end{figure}

For each trajectory $r$, we evaluate the branch-resolved
target fidelity
\[
F_r(t)=|\langle\alpha_0|\psi_r(t)\rangle|^2.
\]
At the conditioning time $t_p=1$, we retain the trajectories
that have already retrieved the addressed branch,
\[\mathcal S(t_p)=\left\{r:F_r(t_p)\geq F_{\mathrm{th}}\right\},
\qquad
F_{\mathrm{th}}=0.9.\]
For this retrieval-conditioned ensemble, the quantity shown
in Fig.~3(a) of the main text is
\[P_{\mathrm{ret}}(t|t_p)=
\frac{1}{|\mathcal S(t_p)|}
\sum_{r\in\mathcal S(t_p)}
\Theta\!\left[F_r(t)-F_{\mathrm{th}}\right].\]
It gives the fraction of trajectories retrieved at $t_p$ that
are above threshold at the specified readout time $t$. This is an
instantaneous readout-time statistic rather than a first-passage survival probability. A trajectory that leaves
the target region and returns before $t$ is therefore counted
as retrieved at that readout time. The definition characterizes
the availability of the stored pattern when the memory is
queried and does not require uninterrupted localization
throughout $(t_p,t)$.
\\
We generate $N_{\mathrm{traj}}=500$ trajectories, of which
$|\mathcal S(t_p)|=384$ satisfy the retrieval criterion at
$t_p=1$. At the final readout time $t_f=10^3$, $147$ of
these trajectories are above threshold, yielding
\[
P_{\mathrm{ret}}(t_f|t_p)
=
\frac{147}{384}
=
0.3828.
\]
The corresponding 95\% Wilson score interval is
$[0.3356,0.4324]$~[\hyperlink{SM-Wilson1927}{4}].
\\~\\
The single threshold fraction does not show how the final
fidelities are distributed within the conditioned ensemble.
Fig.~\ref{fig:jump_stats} therefore displays the complete distribution of
$F_r(t_f)$ for all $384$ trajectories in
$\mathcal S(t_p)$. It shows the ensemble underlying the
reported value of $P_{\mathrm{ret}}(t_f|t_p)$ and separates
the records that remain above the retrieval threshold from
those that have lost the addressed branch by the final
readout.

\section{Linear-gain control}
\label{sec:gain}

To test the channel--manifold compatibility mechanism, we replace the linear
loss jump by
\begin{equation}
 J_1^{(+)}=\sqrt{\gamma_1}\,a^\dagger,
\end{equation}
while retaining
\(J_n=\sqrt{\gamma_n}(a^n-\beta_n)\).  For the loss and gain controls, the corresponding no-click generators differ only by a scalar imaginary shift. Using
\(aa^\dagger=a^\dagger a+\mathbb I\),
\begin{equation*}
H_{\rm eff}^{(+)}
=
H_{\rm eff}^{(-)}
-\frac{i\gamma_1}{2}\mathbb I .
\end{equation*}
Hence, for the same initial state, \(\ket{\widetilde\psi^{(+)}(t)}=e^{-\gamma_1 t/2}\ket{\widetilde\psi^{(-)}(t)}\),
so the normalized no-click states are identical,
\(\ket{\psi_0^{(+)}(t)}=\ket{\psi_0^{(-)}(t)}\),
while their no-click probabilities satisfy
\(P_0^{(+)}(t)=e^{-\gamma_1 t}P_0^{(-)}(t)\).
Thus replacing loss by gain leaves the normalized no-click evolution unchanged, while modifying both the detection statistics and the post-jump state update.
\\~\\
This replacement changes the
unconditional Liouvillian and is therefore a control model, not a different
unraveling of the loss model.  The comparison isolates the action of the
resolved linear jump on the coherent branch.  Linear loss is compatible with
that branch because
\begin{equation}
 \frac{a\ket{\alpha_j}}{\|a\ket{\alpha_j}\|}
 =e^{i\arg\alpha_j}\ket{\alpha_j},
\end{equation}
whereas
\(a^\dagger\ket{\alpha_j}\not\propto\ket{\alpha_j}\).  A gain detection
therefore produces photon-added, noncoherent backaction. For each channel
$\chi\in\{\mathrm{loss},\mathrm{gain}\}$, let
\begin{equation}
\mathcal S_\chi(t_p)
=
\left\{
r:F_r(t_p)\geq F_{\rm th}
\right\}
\end{equation}
denote the corresponding retrieval-conditioned trajectory
set. We denote by $p_\chi(F;t_f)$ the normalized
distribution of the final target fidelities $F_r(t_f)$ over
the trajectories in $\mathcal S_\chi(t_p)$. It satisfies
\begin{equation}
\int_0^1 dF\,p_\chi(F;t_f)=1,
\qquad
P_{\rm ret}^{(\chi)}(t_f\mid t_p)
=
\int_{F_{\rm th}}^1 dF\,p_\chi(F;t_f).
\label{eq:final_fidelity_density}
\end{equation}
Thus the area of the distribution above $F_{\rm th}$ is the
fraction of trajectories retrieved at $t_p$ that remain
above threshold at the final readout.
\begin{figure}[h]
 \centering
 \includegraphics[width=0.96\linewidth]{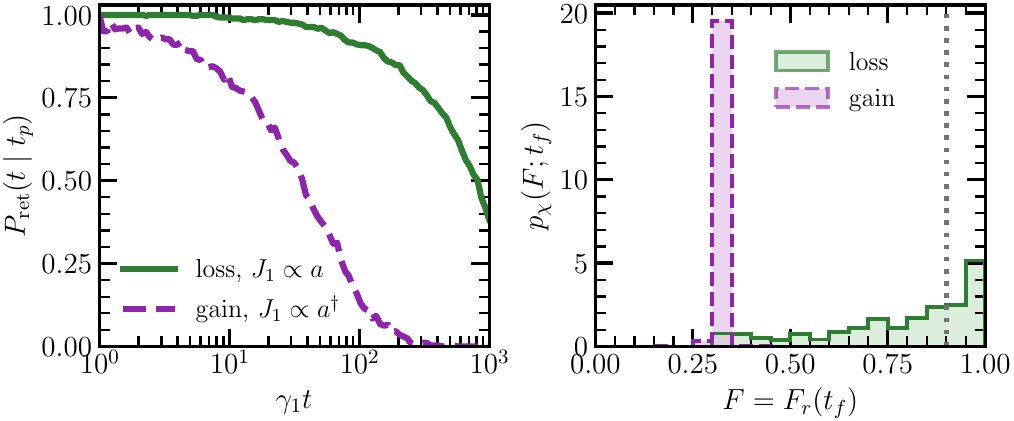}
 \caption{\label{figS4} \textit{Dynamics under linear loss and gain monitoring channels.} Each channel is evaluated over its own retrieval-conditioned set $\mathcal S_\chi(t_p)$ at $t_p=1$, obtained from 500 exact waiting-time trajectories. \textit{Left panel:} Instantaneous retrieval probability. The loss channel
retains a finite retrieved fraction at $t_f=10^3$, whereas
the gain channel fraction vanishes in the simulated
ensemble. \textit{Right panel:} normalized final-fidelity
distributions $p_\chi(F;t_f)$ for the two conditioned
ensembles. The dotted line marks $F_{\rm th}=0.9$; the area
to its right equals
$P_{\rm ret}^{(\chi)}(t_f\mid t_p)$.}
\end{figure}
For the gain control, 368 of 500 trajectories satisfy the retrieval criterion at \(t_p\), but none satisfies it at \(t_f=10^3\), as shown in Fig.~\ref{figS4}.  The 95\% Wilson interval for the final gain-channel fraction is \([0,0.0103]\).  This finite-sample result should be read as an upper bound at the simulated readout time, not as a proof of a strictly zero asymptotic probability. Together with the loss result \(147/384\), it nevertheless shows that long-lived trajectory-level retrieval is strongly enhanced when the frequently resolved jump acts compatibly with the coherent memory structure.

\section*{References}
\begin{list}{}{%
  \setlength{\leftmargin}{2em}%
  \setlength{\labelwidth}{1.6em}%
  \setlength{\labelsep}{0.4em}%
  \setlength{\itemsep}{0.35em}%
  \setlength{\parsep}{0pt}%
}

\item[\hypertarget{SM-WisemanMilburn2010}{[1]}]
H.~M. Wiseman and G.~J. Milburn,
\textit{\href{https://doi.org/10.1017/CBO9780511813948}
{Quantum Measurement and Control}}
(Cambridge University Press, Cambridge, 2010).

\item[\hypertarget{SM-Dalibard1992}{[2]}]
J. Dalibard, Y. Castin, and K. M\o lmer,
Wave-function approach to dissipative processes in quantum optics,
\href{https://doi.org/10.1103/PhysRevLett.68.580}
{Phys. Rev. Lett. \textbf{68}, 580 (1992)}.

\item[\hypertarget{SM-PlenioKnight1998}{[3]}]
M.~B. Plenio and P.~L. Knight,
The quantum-jump approach to dissipative dynamics in quantum optics,
\href{https://doi.org/10.1103/RevModPhys.70.101}
{Rev. Mod. Phys. \textbf{70}, 101 (1998)}.

\item[\hypertarget{SM-Wilson1927}{[4]}]
E.~B. Wilson,
Probable inference, the law of succession, and statistical inference,
\href{https://doi.org/10.1080/01621459.1927.10502953}
{J. Am. Stat. Assoc. \textbf{22}, 209 (1927)}.

\item[\hypertarget{SM-Kessler2012}{[5]}]
E.~M. Kessler,
Generalized Schrieffer--Wolff formalism for dissipative systems,
\href{https://doi.org/10.1103/PhysRevA.86.012126}
{Phys. Rev. A \textbf{86}, 012126 (2012)}.

\item[\hypertarget{SM-ReiterSorensen2012}{[6]}]
F. Reiter and A.~S. S{\o}rensen,
Effective operator formalism for open quantum systems,
\href{https://doi.org/10.1103/PhysRevA.85.032111}
{Phys. Rev. A \textbf{85}, 032111 (2012)}.

\end{list}

\end{document}